\documentclass[range]{ar2e}

\usepackage{epsfig}	
\newcommand	\gtsim	{\lower.5ex\hbox{$\; \buildrel > \over \sim \;$}}
\newcommand     \ltsim  {\lower.5ex\hbox{$\; \buildrel < \over \sim \;$}}

\newcommand     \beq    {\begin{equation}}      
\newcommand     \eeq    {\end{equation}}        

\newcommand     \Angstrom {{\rm \AA}}
\newcommand     \cm     {{\rm \,cm}}
\newcommand     \erg    {{\rm \,erg}}

\newcommand     \g      {{\rm \,g}}

\newcommand     \K      {{\rm \,K}}
\newcommand     \kms    {{\rm km\,s^{-1}}}
\newcommand	\magn	{{\rm mag}}
\newcommand     \micron {{\,\mu{\rm m}}}
\newcommand     \s      {{\rm \,s}}
\newcommand     \yr     {{\rm \,yr}}

\newcommand     \rmH    {{\rm H}}

\newcommand     \sol    {{\odot}}
\newcommand	\etal	{{et al.\ }}

\begin{document}

\jname{Annu. Rev. Astron. Astrophys.}
\jyear{2002}
\jvol{41}
\ARinfo{Preprint astro-ph/...}

\title{Interstellar Dust Grains}

\markboth{Draine}{Interstellar Dust Grains}

\author{B. T. Draine
\affiliation{Princeton University Observatory, Princeton, NJ 08544 USA\\
draine@astro.princeton.edu}}

\begin{keywords}
infrared astronomy, 
interstellar medium, 
interstellar grains, 
light scattering, 
molecular clouds
\end{keywords}

\begin{abstract}
This review surveys the observed properties of interstellar dust grains:
the wavelength-dependent extinction of starlight, including 
absorption features, from
UV to infrared;
optical luminescence; 
infrared emission;
microwave emission;
optical, UV, and X-ray scattering by dust; 
and polarization of starlight and of infrared emission.
The relationship between presolar grains in meteorites and the interstellar
grain population is discussed.
Candidate grain materials and abundance constraints are considered.
A dust model consisting of amorphous silicate grains, graphite grains,
and polycyclic aromatic hydrocarbons is compared with observed emission and
scattering.
Some issues concerning evolution of interstellar dust are discussed.
\end{abstract}

\maketitle

\section{INTRODUCTION}

Dust grains play a central role in the astrophysics of the interstellar
medium, from the thermodynamics and chemistry of the gas,
to the dynamics of star formation.  
In addition, dust shapes the
spectra of galaxies: radiation at short wavelengths is attenuated, and
energy is radiated in the infrared.  It is estimated that 30\% or more
of the energy emitted as starlight in the Universe 
is reradiated by dust in the infrared 
\citep{BFM02}.
Interstellar dust determines what galaxies look like,
how the interstellar medium (ISM) in a galaxy behaves, and the very process of
star formation that creates a visible galaxy.

This review summarizes current knowledge of the abundance,
composition, and sizes of interstellar dust grains, as indicated by
observations of extinction, scattering, and emission from dust,
supplemented by evidence from interstellar gas phase abundances and
presolar grains in meteorites.
The term ``dust grain'' is understood here to extend down to
molecules containing tens of atoms, as there is no discontinuity in the
physics as the particle size decreases from microns to Angstroms.

The distribution of gas and dust in the Galaxy -- the structure 
of the interstellar medium -- is not discussed.

The astrophysics of interstellar dust is not covered in this review.
This includes the physical optics of small particles, charging of dust grains, 
heating and cooling of dust grains, 
chemistry on dust grain surfaces, forces and torques
on dust grains, sputtering and shattering of dust grains, 
and alignment of dust with the magnetic field; an introduction to these
topics can be found elsewhere 
\citep{Kru02,Dr03a}.

The review does not attempt to survey or compare different models proposed for 
interstellar dust.  One dust model is introduced to illustrate
modeling of absorption, scattering, and infrared emission from interstellar
dust.

It is not possible to cite all of the important papers in this area;
only a few articles are cited in connection with each topic.
The reader is encouraged to also consult prior reviews by
\citet{SM79},
\citet{Ma90,Ma93,Ma00},
\citet{Dr95}, 
\citet{Wi00a}, 
\citet{Wi00b},
\citet{Vos02},
and the book by \citet{Whi03}.
\citet{DH95} discuss the metamorphosis of interstellar dust.


\section{EXTINCTION}

The existence of interstellar dust was first inferred from obscuration,
or ``extinction'', of starlight 
\citep{Tr30}.
Much of our knowledge of interstellar dust continues to be
based on studies of the
wavelength-dependence of this attenuation, often referred to
as ``reddening'' because of the tendency for the extinction
to be greater in the blue than in the red.
The wavelength-dependence strongly constrains the grain size 
distribution, and spectral features (\S\ref{sec:spectroscopy}) reveal
the chemical composition.

The extinction is most reliably 
determined using the ``pair method'' -- comparing spectrophotometry of two 
stars
of the same spectral class; if one star has negligible foreground dust
while the second star is heavily reddened, comparison of the two spectra,
together with the assumption that the dust extinction goes to zero at
very long wavelength, allows one to determine the 
extinction $A_\lambda\equiv 2.5\log_{10}(F_\lambda^0/F_\lambda)$
as a function of wavelength $\lambda$,
where $F_\lambda$ is the observed flux and $F_\lambda^0$ is the flux
in the absence of extinction.
The ``pair method'' has been used to measure extinction curves for many
sightlines, in many cases over a range of wavelengths extending from
the near-infrared to the vacuum UV.

\subsection{Milky Way Dust}

\subsubsection{OPTICAL-UV EXTINCTION CURVES
	\label{sec:optical-uv}}
The dimensionless quantity $R_V\equiv A_V/(A_B-A_V)$ is
a common measure of the slope of the extinction curve in the optical
region.
Very large grains would produce gray extinction with $R_V\rightarrow\infty$.
Rayleigh scattering ($A_\lambda\propto\lambda^{-4}$) 
would produce very steep extinction with
$R_V\approx 1.2$.
$R_V$ is known to vary from
one sightline to another, from values as low as 2.1 
(toward HD~210121) \citep{WF92} 
to values as large as 5.6-5.8 
(toward HD~36982) \citep{CCM89,Fi99}.
\citet{CCM89} showed that normalized
extinction curves $A_\lambda/A_{I}$ (using the $I$ 
band extinction to normalize)
could be approximated by a seven-parameter function of wavelength $\lambda$:
\begin{equation}
A_\lambda/A_{I} \approx f(\lambda;R_V,C_1,C_2,C_3,C_4,\lambda_0,\gamma) ~~.
\label{eq:CCM}
\end{equation}
At wavelengths $\lambda > 3030\Angstrom$, the function $f(\lambda)$ 
depends only on
$\lambda$ and the single parameter $R_V$.
The parameters $C_3$, $\lambda_0$, and $\gamma$ determine the
strength and shape of the 2175\AA\ ``bump'', and the coefficients
$C_1$, $C_2$, and $C_4$ determine the slope and curvature of the continuous
extinction at $\lambda < 3030\Angstrom$.

\citet{CCM89} show that if the single quantity $R_V$ is known, 
it is possible to
estimate the values of the other six parameters ($C_{1-4}$, $\lambda_0$,
$\gamma$) so that the optical-UV extinction can be approximated by a
one-parameter family of curves.  However, if the UV
extinction has been measured, an improved fit to the observations can
be obtained by fitting $C_{1-4}$, $\lambda_0$ and $\gamma$ to the measured
extinction.

\begin{figure}[ht]

\begin{center}
\epsfig{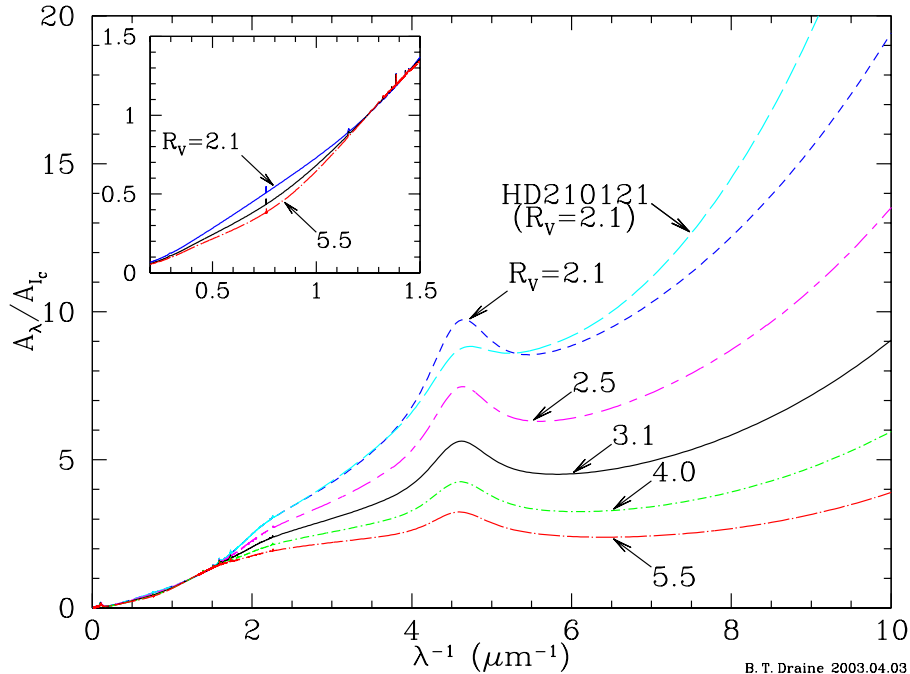}
\vspace*{-0.3cm}
\end{center}

%
%
%
\caption{\footnotesize
	\label{fig:ext}
	Extinction curves from prescription of 
	Fitzpatrick (1999),
	with
	diffuse interstellar bands (DIBs)
	added as described in
	\S\ref{sec:DIBs}.
	The DIBs are barely visible on this plot.
	}
\end{figure}

Empirical extinction curves in diffuse clouds
show relatively little
variation in the infrared, and for wavelengths
$0.7\micron\ltsim\lambda\ltsim 8\micron$ the 
function (\ref{eq:CCM}) appears to be approximately
``universal''
(i.e., independent of $R_V$)
in diffuse clouds
[in dense clouds grains acquire 
``ice'' mantles (\S\ref{sec:ices}) which alter the extinction].
\citet{Fi99}
pays careful attention to the effects of
finite-width photometric bandpasses, and
gives modified formulae 
which appear to improve the overall fit to observations of
ice-free dust.
The \citet{CCM89} and \citet{Fi99} fits for $R_V=3.1$ are compared 
in \S\ref{sec:scattering}
(Table \ref{tab:model_IRoptuv}).

Figure \ref{fig:ext} shows extinction curves calculated using the
F99 parametrization for $R_V=2.1$, 2.5, 3.1, 4.0, and 5.5.
The coefficients in the \citet{CCM89} or \citet{Fi99} fitting formulae can be
adjusted to improve the fit to specific sightlines; such a fit is shown
for the extreme case of HD~210121, showing that the UV extinction
can differ significantly from the average behavior for the same value
of $R_V$.

``Pair method'' determinations of the reddening law
for many sightlines indicate that $R_V\approx3.1$ for 
the ``average'' extinction law for diffuse regions in the local Milky Way
\citep{SM79,CCM89}.
Sightlines intersecting clouds with larger extinction per cloud tend to
have larger values of $R_V$; the larger $R_V$ values may indicate
grain growth by accretion and coagulation.

Another approach to determining the wavelength-dependent extinction is
to use star counts or galaxy counts as a function of apparent brightness.
\citet{SG99}
use UBVRI photometry 
of Galactic field stars
to determine the extinction law; for the four 
high-latitude clouds they studied,
they conclude that $R_V\ltsim 2$,
well below the value $R_V\approx3.1$ which
is widely considered to be ``average'',
and comparable to the most extreme values of
$R_V$ ever found in studies of individual stars ($R_V=2.1$ for HD~210121).
Further study is needed to reconcile this apparent conflict.

\citet{Ud02}
uses $V$ and $I$ photometry of ``red clump giants'' 
to study the reddening law.
Toward the LMC he infers $R_V\approx 3.1$, but toward Galactic bulge regions
he finds 
$R_V\approx 1.8 - 2.5$.

Sloan Digital Sky Survey (SDSS) photometry 
\citep{LI03}
can be used to study the reddening toward
stars at the ``blue tip'' of the main sequence 
\citep{FS03}.
The observed reddening appears to be consistent with
$R_V\approx3.1$.
Galaxy surface brightnesses can also be used;
SDSS galaxy photometry
is consistent
with $R_V=3.1$ 
\citep{SFB03}.

\subsubsection{EXTINCTION PER H}
Using H Lyman-$\alpha$ and absorption lines of H$_2$ to determine the
total
H column density $N_{\rm H}$, 
\citet{BSD78}
found
\beq
N_{\rm H}/(A_B-A_V)= 5.8\times10^{21}\cm^{-2} \magn^{-1} ~~~,
\eeq
\beq
\label{eq:AV_per_H}
N_{\rm H}/A_V \approx 1.87\times10^{21}\cm^{-2} \magn^{-1}
~~~{\rm for}~R_V=3.1 ~~~,
\eeq
to be representative of dust in diffuse regions.
For $R_V=3.1$, the F99 reddening fit gives $A_{I_C}/A_V=0.554$ for
Cousins $I$ band ($\lambda=0.802\micron$), thus
\beq
A_{I_C}/N_{\rm H}\approx 2.96\times10^{-22}\magn\cm^{2}
~~~{\rm for}~R_V=3.1 ~~.
\label{eq:AI_per_H}
\eeq
If the extinction curve is really universal for $\lambda\gtsim0.7\micron$
we might expect eq.\ (\ref{eq:AI_per_H}) to apply independent of $R_V$.

\begin{figure}[ht]
\begin{center}
\epsfig{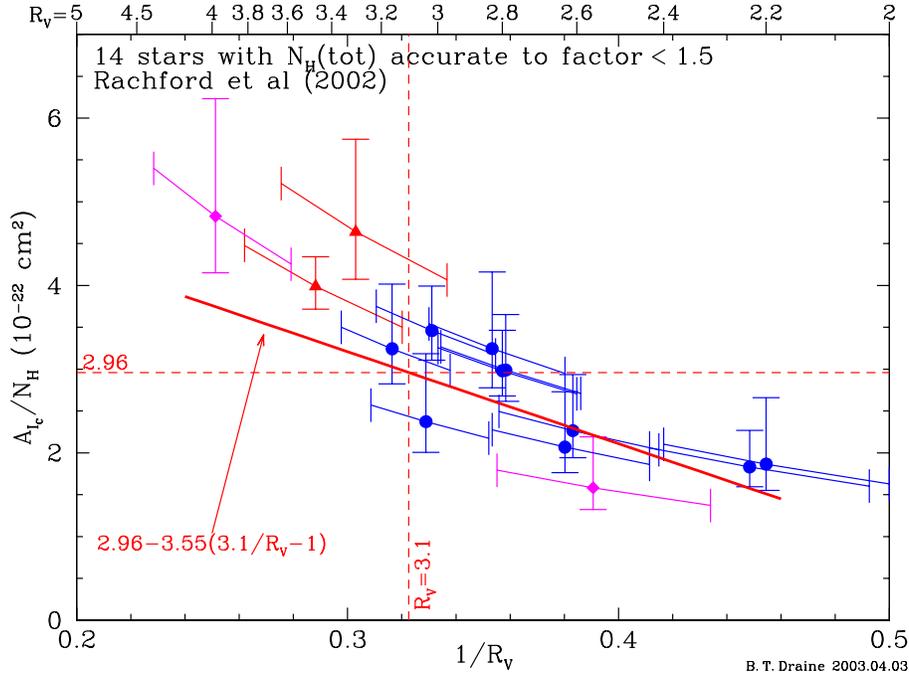}
\vspace*{-0.3cm}
\end{center}
%
%
%
\caption{\footnotesize
	$A_I/N_{\rm H}$ for 14 sightlines through translucent clouds 
	\citep{RST02}, as a function
	of $1/R_V$, with
	$R_V$ determined by IR photometry (circles, $R_V\pm0.2$);
	polarization $\lambda_{\rm max}$ (triangles, $R_V\pm10\%$);
	or UV extinction curve (diamonds, $R_V\pm10\%$).
	Vertical error bars show 1-$\sigma$ uncertainty due to
	errors in $N_{\rm H}$; tilted error bars show effects
	of errors in $R_V$.
	Least-squares fit [eq.\ \ref{eq:A_I/N_H})] is shown.
	\label{fig:extperH}}
\end{figure}

\citet{RST02}
report H and H$_2$ column densities
for 16 sightlines through diffuse and ``translucent'' clouds.
Figure \ref{fig:extperH} shows $A_{I_C}/N_{\rm H}$ versus $R_V^{-1}$
for the 14 sightlines with
$N_{\rm H}$ known to better than a factor 1.5 .
The $I_C$ band extinction $A_{I_C}$ has been estimated from $A_B-A_V$ and $R_V$
using the F99 $R_V$-dependent reddening law.

\begin{figure}[ht]
\begin{center}
\epsfig{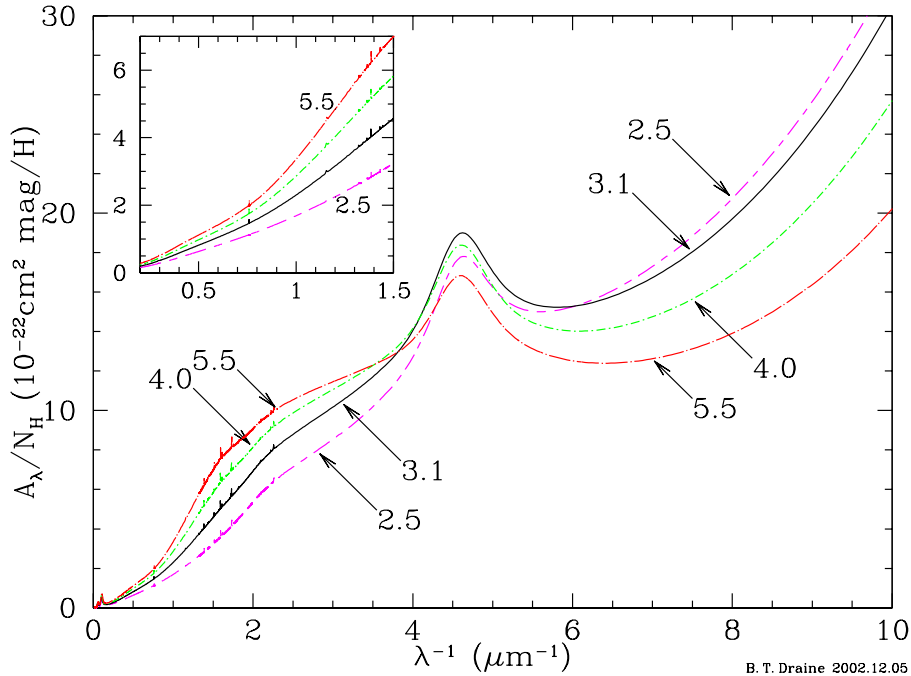}
\vspace*{-0.3cm}
\end{center}
%
%
%
\caption{\footnotesize
	$A_\lambda/N_{\rm H}$ for different values of $R_V$,
	using F99 extinction curve fit and $A_I/N_{\rm H}$ from
	eq.\ (\ref{eq:A_I/N_H}).
	\label{fig:extcurvperH}}
\end{figure}

Figure \ref{fig:extperH} appears to show that $A_{I_C}/N_{\rm H}$ is {\it not}
universal for local Milky Way dust: the $I_C$ band extinction
per H 
deviates from the ratio (\ref{eq:AI_per_H})
when $R_V$ deviates from $3.1$, in the sense that $A_{I_C}/N_{\rm H}$
increases when $R_V$ is larger (larger grains).
We note that the trend in $A_{I_C}/N_{\rm H}$ could be in part the result of
errors in determination of $R_V$,
(i.e., errors in estimating $A_V$,
since $E(B-V)$ is probably accurate to better than 10\% for these stars),
but the data appear to show a real trend.
A least-squares fit, constrained to pass through (\ref{eq:AI_per_H}), gives
\beq
A_{I_C}/N_{\rm H} \approx \left[2.96-
3.55\left(3.1/R_V-1\right)\right]\times10^{-22}\magn\cm^{2}
~~~.
\label{eq:A_I/N_H}
\eeq

Figure \ref{fig:extcurvperH} shows
$A_\lambda/N_{\rm H}$, the extinction per H, 
estimated from the F99 extinction fits and eq.\ (\ref{eq:A_I/N_H}).
If the F99 extinction fits and eq.\ (\ref{eq:A_I/N_H}) are both correct,
the extinction per H {\it increases}
for $\lambda \gtsim 0.3\micron$ when $R_V$ increases.
Presumably this is due to increased scattering by the
grains as the grain size increases.

\subsubsection{INFRARED EXTINCTION
	 \label{sec:IR}}

Between $\sim 0.9\micron$ and $\sim5\micron$ 
the continuous extinction curve can be approximated by a power-law,
$A_\lambda \propto \lambda^{-\beta}$,
with $\beta\approx 1.61$ \citep{RL85},
1.70 \citep{Wh88}, 1.75 \citep{Dr89b},
$\sim$1.8 \citep{MW90},
$\sim$1.8 \citep{WMF93},
or 1.70 \citep{BTR99,RBD00}.

\begin{figure}[ht]
\begin{center}
\epsfig{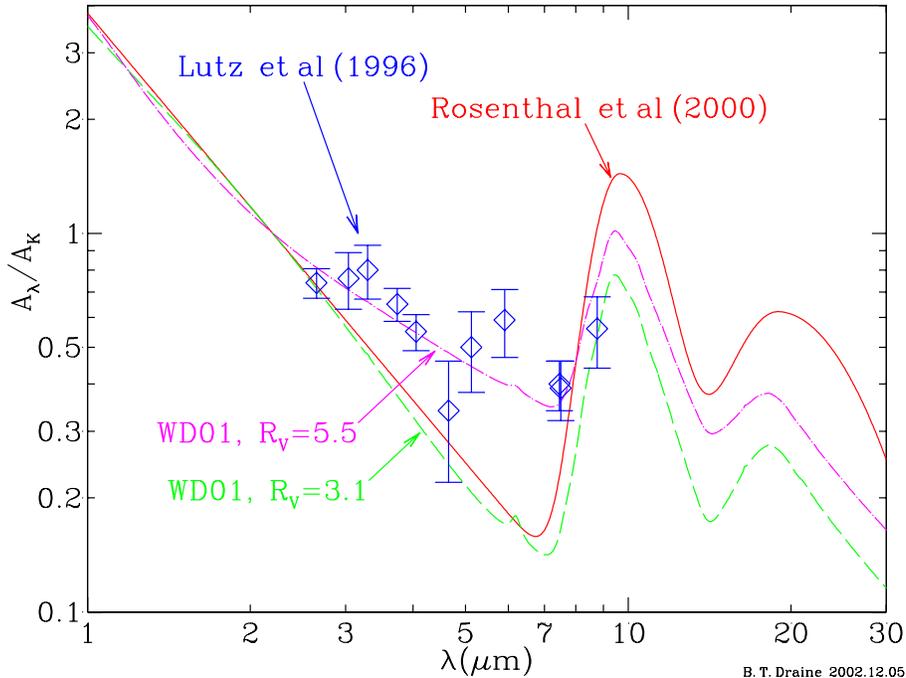}
\vspace*{-0.3cm}
\end{center}
%
%
%
\caption{\footnotesize
	\label{fig:ext_ir}
	Infrared extinction, relative to extinction at K (2.2$\micron$),
	inferred by 
	\citet{RBD00}
	for dust in the OMC-1 molecular cloud,
	and by \citet{LFG96}
	for dust toward Sgr A$^*$, and as
	calculated for the 
	\citet{WD01a}
	grain models
	for $R_V=3.1$ (60ppm C in PAH) and $R_V=5.5$ (30ppm C in PAH).
	The 6.2$\micron$ absorption feature in the \citet{WD01a} 
	grain models is
	due to PAHs.
	}
\end{figure}

The extinction in the 5--8$\micron$ region is controversial 
(see Figure \ref{fig:ext_ir}).
\citet{Dr89b}
concluded that the observational evidence was consistent
with $A_\lambda \propto \lambda^{-1.75}$ out to $\lambda \approx 6\micron$.
Using ISO observations of H recombination lines, 
\citet{LFG96}
find that the extinction toward Sgr A$^*$
does not decline with increasing $\lambda$
in the 4--8$\micron$ region.
However, studies 
using H$_2$ rovibrational lines in the Orion molecular cloud, 
\citep{BTR99,RBD00}
find that the extinction continues to 
decline with increasing $\lambda$ to a minimum
at $\sim$6.5$\micron$.
The wavelength dependence of the continuous dust extinction
in the Orion molecular cloud could conceivably differ from that
on the sightline to Sgr A$^*$; alternatively, perhaps
the H recombination line intensity ratios differ from the
``case B'' recombination assumed by Lutz \etal
It is interesting to see in Figure 
\ref{fig:ext_ir} that the \citet{WD01a} grain
model for $R_V=5.5$ is in rough agreement with the Lutz \etal results
toward the galactic center.
Further observational study of the 5--8$\micron$ extinction is needed.

In addition to this continuous extinction, there are strong absorption
features at 9.7$\micron$ and 18$\micron$ due to silicates 
(\S\ref{sec:silicate}), a hydrocarbon feature at 3.4$\micron$
(\S\ref{sec:CH}), and PAH absorption at 6.2$\micron$ (\S\ref{sec:PAHs}).
The extinction in molecular clouds also shows additional bands
due to ice mantles (\S\ref{sec:ices}).

\subsection{Dust in Other Galaxies}

Dust in the Milky Way varies from one sightline to another.
It is obviously of great interest to study the dust in other galaxies,
both in order to correct observations for the extinction by the dust,
but also to inform our understanding of dust astrophysics.

Our knowledge of the extinction law for dust in other galaxies is
greatest for the Large and Small Magellanic Clouds (LMC and SMC), 
where we can study the extinction for individual stars, but some progress
has been made in determining the wavelength-dependence of extinction
in more distant galaxies.

\subsubsection{MAGELLANIC CLOUDS}
Not surprisingly, there are regional variations within the LMC and SMC.
Stars more than $\sim$500 pc away from the 30 Doradus region of the LMC
have $R_V\approx 3.2$ and an extinction law that appears to be quite
similar to the Milky Way diffuse cloud extinction law 
\citep{Fi86,MCG99}.
The reddening per H atom 
$E(B-V)/N_{\rm H}\approx 4.5\times10^{-23}\cm^{2}/{\rm H}$
\citep{Ko82,Fi85},
approximately 40\% of the Milky Way value.
This is approximately consistent with the LMC metallicity: LMC H~II regions 
have Ne/H$\approx(3.5\pm1.4)\times10^{-5}$ 
\citep{KD98},
$\sim$30\% of the
solar Ne/H 
($=1.2\times10^{-4}$) \citep{GS98}.
Stars in or near 30 Dor have extinction curves which have a weakened
2175\AA\ feature 
\citep{Fi86,MCG99}.

In the SMC, stars in the ``bar'' region have extinction
curves which appear to lack the 2175\AA\ feature.
The reddening per H is only 
$E(B-V)/N_{\rm H}=2.2\times10^{-23}\cm^{2}/{\rm H}$, 
$\sim$13\% of the Milky Way value, again consistent with the
SMC metallicity: SMC H~II regions have
Ne/H$\approx(1.6\pm0.2)\times10^{-5}$ 
\citep{Du84,KD98},
or 14\% of the solar Ne/H,
and the gas-phase C/H is only 6\% of solar.

\citet{WD01a}
are able to reproduce the
observed extinction laws in the LMC and SMC using appropriate mixtures
of carbonaceous and silicate grains.  The sightlines in the SMC bar
which lack the 2175\AA\ extinction feature can be reproduced by
models which lack carbonaceous grains with radii $a\ltsim 0.02\micron$.

\subsubsection{M31}
\citet{BCB96}
present UV
extinction curves toward selected bright stars in M31, finding an
extinction law similar to the average Milky Way extinction curve, but
with the 2175\AA\ feature possibly somewhat weaker than in the Milky Way.

\subsubsection{OTHER GALAXIES}
There are several different approaches to determining the extinction curve
for dust in galaxies where individual stars cannot be resolved.

The dust extinction law can be determined when a foreground galaxy
overlaps a background galaxy.
\citet{BQP97}
find the extinction curve for dust in the spiral arm of NGC 2207 to resemble
the Milky Way extinction law for $R_V=5.0$; the interarm dust appears
to be even grayer.  However, the flatness of the extinction curve could
be due in part to unresolved optically-thick dust patches in the foreground
galaxy.

\citet{KW01a}
find the extinction law in the spiral galaxy
AM 1316-241 to be close to the Milky Way mean extinction,
with $R_V=3.4\pm0.2$; for the spiral galaxy AM 0500-620 the dust extinction
law appears to be somewhat steeper than the Milky Way average, with
$R_V\approx 2.5\pm0.4$.
\citet{KW01b}
find the extinction law in the Sc galaxy NGC 3314A to
be close to the Milky Way mean, with $R_V=3.5\pm0.2$ over 
galactocentric radii 1.6--3.8 kpc, with no evidence for any radial trend in
reddening law.

From observations of the overall emission spectrum
\citet{Ca01}
infers the internal extinction by dust in starburst galaxies.
The inferred reddening law is relatively gray compared to
the Galactic reddening law, and shows no evidence of a 2175\AA\ feature.
It is not clear to what extent the flatness of the apparent 
reddening law
may be due to the effects of radiative transfer in optically-thick
distributions of stars and dust.

Searches have been made for the 2175\AA\ extinction feature in QSO spectra
\citep{MP74}.
\citet{Ma97}
reports a statistical detection in a sample of 92 QSOs, but
\citet{PCG00} 
argue that the detection was not statistically significant.

Gravitationally-lensed QSOs with multiple images can be used to
determine the extinction law if the intrinsic spectrum is assumed to
be time-independent.
\citet{FIK99}  
determine extinction laws in galaxies out to $z=1.01$,
with $R_V$ estimates from 1.5 to 7.2.
\citet{THB00}
find the extinction curve in a galaxy at $z=0.44$ to
have $1.3 \ltsim R_V\ltsim 2.0$.
\citet{MMM02}
find that the dust in a lensing galaxy at $z=0.83$ is
consistent with a standard Milky Way extinction law for
$R_V=2.1\pm0.9$, including an extinction bump at 2175\AA.

\section{SPECTROSCOPY OF DUST\label{sec:spectroscopy}}

The composition of interstellar dust remains controversial.
While meteorites (\S\ref{sec:meteor}) 
provide us with genuine specimens of interstellar
grains for examination, these are subject to severe selection effects,
and cannot be considered representative of
interstellar grains.
Our only direct information on the composition of interstellar dust
comes from spectral features in extinction, scattering, or emission.

\subsection{2175\AA\ Feature: Aromatic C?\label{sec:2175}}

By far the strongest spectral feature is the broad ``bump'' in the
extinction curve centered at $\sim$2175\AA\ (see Figure \ref{fig:ext}).

\citet{FM86}
showed that the observed 2175\AA\ feature can be accurately fit with a
Drude profile
\beq
\Delta C_{\rm ext}(\lambda) = 
\frac{C_0 \gamma^2}{(\lambda/\lambda_0-\lambda_0/\lambda)^2+
\gamma^2} ~~,
\eeq
peaking at $\lambda_0=2175\Angstrom$ (or $\lambda_0^{-1}=4.60\micron^{-1}$)
and with a broadening parameter 
$\gamma=0.216$
(corresponding to
FWHM $\gamma\lambda_0=469\Angstrom$,
or $\gamma\lambda_0^{-1}=0.992\micron^{-1}$).
The strength of the absorption is
such that on an average diffuse cloud sightline in the Milky Way,
the 2175\AA\ profile corresponds to an oscillator strength per H nucleon
$n_{\rm X}f_{\rm X}/n_\rmH \approx 9.3\times10^{-6}$ 
\citep{Dr89a}.
Since we expect the oscillator strength per molecule $f_{\rm X}\ltsim 0.5$,
it can be concluded that the carrier molecule must contain
one or more elements from \{C,O,Mg,Si,Fe\}, as these are the only 
condensible elements
with sufficiently high abundances
$n_{\rm X}/n_\rmH\gtsim4\times10^{-5}$ in the ISM.

\citet{FM86}
show that the width of the
feature varies considerably from one sightline to another.
The FWHM has an average value $0.992\micron^{-1}$, but the distribution
has a $\pm2\sigma$ variation of $\pm12\%$.
This variability in FWHM is in contrast to the fact that
the central wavelength $\lambda_0$ is nearly invariant from one
sightline to another: although variations are detected, $\lambda_0$ shows
$\pm2\sigma$ variations of only $\pm0.46\%$.

Immediately following discovery of the 2175\AA\ feature by 
\citet{St65}, \citet{SD65}
pointed out that small graphite particles
would produce absorption very similar to
the observed feature.
Although non-carbonaceous carriers have been proposed 
\citep[e.g., OH$^-$ on small silicate grains:][]{SD87},
it now seems likely that some form of graphitic carbon is responsible
for the observed feature.

The C atoms in graphite are bound to one another in hexagonal sheets.
Each C atom has 4 valence electrons ($2s^22p^2$); 
3 of these are in trigonal $sp^2$ or $\sigma$ orbitals;
the remaining electron is in a delocalized $\pi$ orbital, which is
shared among the C-C bonds.
The individual sheets of carbon atoms are held parallel to one another
only by weak van der Waals forces.
Absorption of a 
photon can excite a $\pi$ electron to an excited
orbital ($\pi^*$); this $\pi\rightarrow\pi^*$ 
transition is responsible for the absorption
feature peaking at $\sim2175\Angstrom$.
Using the dielectric function of graphite 
\citep{DL84},
one finds that randomly-oriented small graphite spheres would have
an oscillator strength $f=0.16$ per C atom 
\citep{Dr89a}.
Thus the observed 2175\AA\ extinction feature would require
C/H=$5.8\times10^{-5}$ in small graphite spheres, compatible with
interstellar abundance constraints.

Although graphite is an attractive candidate, the graphite hypothesis
does not appear to have a natural way to accomodate
the observed variations in
FWHM of the profile while at the same time holding $\lambda_0$ nearly
constant: variations in graphite grain shape and size 
produce variations in FWHM,
but these are accompanied by changes in $\lambda_0$
\citep{DM93}.

\citet{MCB98}
propose that the 2175\AA\ feature is due to UV-processed
HAC particles, but do not appear to have actually measured in the lab
a profile matching the interstellar 2175\AA\ feature.

\citet{WKK99}
propose that the 2175\AA\ feature is due to onion-like
hyperfullerene carbon particles.  While these particles do have an
absorption peak near 2200\AA, the feature appears to be broader than 
the interstellar profile.

The carbon atom skeleton of polycyclic aromatic hydrocarbon (PAH) molecules
is very similar to a portion of a graphite sheet, with similar
electronic wavefunctions.
It is therefore not surprising that PAH molecules
generally have strong $\pi\rightarrow\pi^*$
absorption in the 2000--2500\AA\ region.
Thus large PAH
molecules are candidates to be the carrier of the interstellar 2175\AA\
feature -- this is a natural extension of the graphite hypothesis.

The grain model of
\citet{WD01a}
and
\citet{LD01b}
has C/H=$6.0\times10^{-5}$ in PAH molecules or clusters 
containing from 20 to $10^5$ C atoms -- this population 
of PAH molecules is required to reproduce the
observed infrared emission
(see \S\ref{sec:IRemission}).
Absorption profiles are not known for PAH molecules of the sizes
characteristic of the interstellar PAHs, but one would expect a similar
$\pi\rightarrow\pi^*$
oscillator strength per C as for graphite, or $f\approx0.16$.
Thus we see that in this grain model, the 2175\AA\ extinction feature
is expected to be primarily -- perhaps entirely -- due to large PAH molecules.

In this interpretation, the observed band profile would 
be due to a mixture of PAHs,
and the observed variations in FWHM (and small variations in $\lambda_0$)
would result from differences in the
PAH mix from one sightline to another.

\subsection{Silicate Features\label{sec:silicate}}

The infrared extinction includes
a strong absorption feature peaking at $\sim$9.7$\micron$.
Silicate minerals generally have strong absorption resonances near 10$\micron$
due to the Si-O stretching mode, and it seems virtually certain that the
interstellar 9.7$\micron$ feature is due to absorption by interstellar
silicate material.
This conclusion is strengthened by the fact that a 10$\micron$ emission
feature is observed in outflows from cool oxygen-rich stars (which would
be expected to condense silicate dust) but not in the outflows from
carbon-rich stars (where silicates do not form, because all of the oxygen
is locked up in CO).
There is also a broad feature at 18$\micron$ which is presumed to be
the O-Si-O bending mode in silicates (McCarthy \etal 1980; Smith \etal 2000).

The Trapezium 
\citep{GFM75}
emission implies $\tau(\lambda)$ with 
FWHM$\approx$3.45$\micron$, whereas $\mu$~Cep 
\citep{RSF75}
has FWHM=2.3$\micron$, so the silicate properties are not universal.
On sightlines dominated by relatively 
diffuse clouds (e.g., toward the B5 hypergiant
Cyg OB2-12 or distant WC stars)
the 9.7$\micron$ absorption appears to be
better fit by the narrower $\mu$~Cep profile
\citep{RA84,BAW98}.
The silicates in molecular clouds, however, seem better fit by
the broader Trapezium profile 
\citep{BAW98}.

The strength of the 9.7$\micron$ feature relative to $A_V$ has been
measured toward Cyg OB2-12 and
toward WC stars.
The studies in Table \ref{tab:AV/tausil} together indicate
$A_V/\Delta\tau_{9.7}=18.5\pm2.0$, where the uncertainty includes
a subjective allowance for systematic errors in the model-fitting.

\begin{table}[h]
\caption{Silicate 9.7$\micron$ Feature Strength}
\label{tab:AV/tausil}
{\footnotesize
\begin{center}
\begin{tabular}{@{}l
	@{\hspace*{0.5em}} c
	@{\hspace*{0.5em}} c
	@{\hspace*{0.5em}} c 
	@{\hspace*{0.5em}} l
	@{\hspace*{0.7em}} l
	@{\hspace*{0.5em}} c 
	@{\hspace*{0.5em}} l
	@{}}
\toprule
Sightline&$l$&$b$&d&$A_V$&$\Delta\tau_{9.7}$	& $A_V/\Delta\tau_{9.7}$	& ref\\
&($^\circ$)&($^\circ$)&(kpc)&(mag)&\\
\hline
Gal. center&0&0&8.5	& $34\pm4$&$3.6\pm0.3$&$9\pm1$	& \citet{RA85}\\
Cyg~OB2-12&80.10&0.83&1.7	& $10$	  &$0.58$ & 17.2& \citet{RA84}\\
``		&&&& $10.2$&$0.54$&$18.9$	& \citet{WBG97}\\
``		&&&& $10.2$&$0.64$&$15.9$	& \citet{BAW98}\\
WR98A&358.13&-0.03&1.90		& $12.5$&$0.62$&$20.2$	& \citet{vdH96}\\
``		&&&& $11.2$&$0.64$&$17.4$	&\citet{SHW98}\\
WR112&12.14&-1.19&4.15		& $13$	&$0.61$&$21.3$	&\citet{RA84}\\
``		&&&& $11.9$&$0.64$&$18.6$	&\citet{vdH96}\\
``		&&&& $12.0$&$0.56$&$21.4$	&\citet{SHW98}\\
WR118&21.80&-0.22&3.13		& $13.3$&$0.69$&$19.3$	&\citet{RA84}\\	
``		&&&& $12.6$&$0.65$&$19.4$	&\citet{vdH96}\\
``		&&&& $12.8$&$0.71$&$18.0$	&\citet{SHW98}\\
\hline
\multicolumn{4}{c}{Local Diffuse ISM} &&& $18.5 \pm2$ & overall\\
\botrule
\end{tabular}\\
\end{center}
}
\end{table}

The interstellar $9.7\micron$ feature is broad and relatively featureless,
as opposed to absorption profiles measured in the laboratory for
crystalline silicates, which show considerable structure which can
be used to identify the precise mineral.
The absence of substructure in the interstellar profile is believed to
indicate that the interstellar silicates are largely amorphous rather
than crystalline.
Amorphous or glassy 
silicates can be produced in the laboratory by ion bombardment
of initially crystalline material 
\citep{KH79},
formation in smokes 
\citep{Da79},
rapid quenching of a melt
\citep{JMB94}
or deposition following evaporation
\citep{KT92,SBB95,SDu96}.
The laboratory absorption profiles are in some cases quite similar
to the interstellar profile,
supporting the view that interstellar silicates are amorphous.

The observed strength per H nucleon 
of the interstellar silicate band 
appears to require that close to 100\% of the solar abundance
of Si, Fe, and Mg be condensed into amorphous grains.
\citet{Ma98}
has argued that the observed strength can only be reproduced
if silicate grains are ``fluffy'', with $\ge$25\% vacuum, as in the
grain model of 
\citet{MW89}.

The observed 10$\micron$ profile appears to be 
consistent with amorphous material
with a composition in the olivine family, Mg$_{2x}$Fe$_{2-2x}$SiO$_4$.
Since Mg and Fe are of approximately equal abundance in the Sun,
it would be reasonable to have $x\approx0.5$.
\citet{LD01a} conclude that
spectrophotometry of the 9.7$\micron$ feature in extinction toward Cyg OB2-12
\citep{BAW98}
limits the fraction of the Si in crystalline silicates to be
$\ltsim5\%$.
\citet{DJD99}
find that for two protostellar objects, at most 2\%
of the silicates could be crystalline.

However, \citet{BA02} argue that the observed 10$\micron$ profile toward
Cyg OB2-12 is consistent with a mixture of amorphous silicates (40\% by mass)
and 8 crystalline pyroxenes (60\% by mass);
each of the crystalline species in their mix contributes
$\sim$7.5\% by mass, and the fine structure in the 10$\micron$ profile
appears to be consistent with the observations.
However, this interpretation raises various concerns:
\begin{itemize}
\item The lab measurements
      were for powders in a KBr matrix; spectra in vacuo would be different.
	\citet{BA02} state that the ``KBr shift'' is minimal, but it
	would be expected to be significant near these strong resonances.
\item A mix of crystalline grains that gives a smooth absorption profile
      will not give a smooth emission profile unless the different crystalline
      materials are heated to the same temperature.
\item The different crystalline types have distinct far-infrared modes.
      The smoothness of the far-infrared spectra of star-forming regions
      places upper limits on crystalline abundances.  
      For example, the proposed mixture has $\sim$7.5\% of the Si atoms in 
      diopside, but the observations of \citet{OO03} (see below)
      limit diopside to $\ltsim2\%$ of interstellar Si.
\item It is not apparent that the proposed mixture would be consistent with
	the 18$\micron$ extinction profile.
\item The polarization profile of grains near strong resonances is sensitive
      to the dielectric function.  It is not apparent that the proposed mixture
	would be consistent with spectropolarimetric constraints (see below).
\end{itemize}
Further study of these points is urgently required.
In the text below it is assumed that interstellar silicates are
predominantly amorphous, but the reader should keep in mind that a significant
crystalline fraction may be possible.

The 9.7 and 18$\micron$ silicate features can be observed in
polarization, either in absorption or in emission.
For strong resonances (such as the two silicate features),
the polarization profile differs from the absorption profile, so that
spectropolarimetry can constrain
the dielectric function of the grain material 
\citep{Ma75}.
The observed intensity and polarization across the 9.7$\micron$ feature
appear to be approximately 
consistent with the ``astronomical silicate'' dielectric
function estimated by 
\citet{DL84},
but the observed 20$\micron$ polarization
is stronger than expected for that dielectric function
\citep{ASR89,SWA00}.

\citet{OO03}
report detection of an emission feature at 65$\micron$
which may be due to diopside CaMgSi$_2$O$_6$,
containing 5-10\% of interstellar Ca 
(corresponding to 0.35-0.7\% of interstellar Si); 
if confirmed, this would be the
first evidence for crystalline silicates in the ISM.
While crystalline silicates are not abundant in the ISM,
there is evidence from distinctive emission features
for their presence in dust disks
around main-sequence stars 
\citep[see][for a summary]{Ar00},
young stellar objects and evolved stars 
\citep[see][for a summary]{WMW00},
and in some solar-system comets 
\citep[see][for a summary]{Ha99};
see \S\ref{sec:cand_silicate} below.
Nevertheless,
even in these systems most of the silicate material remains amorphous.

\begin{figure}[htb]
\begin{center}
\epsfig{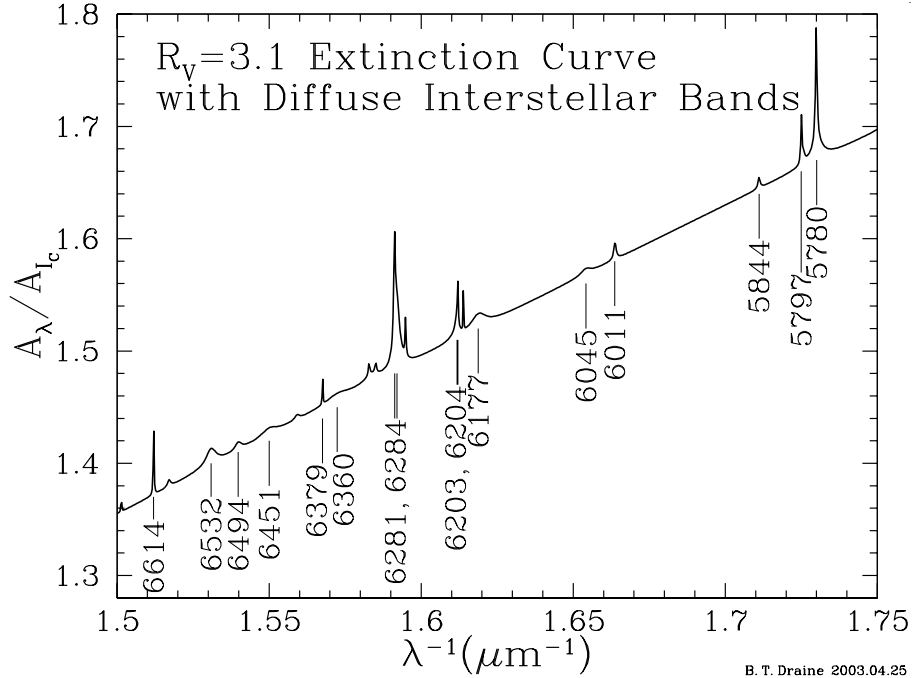}
\vspace*{-0.3cm}
\end{center}
%
%
%
\caption{\footnotesize
	\label{fig:dibs}
	Diffuse interstellar bands in the 5700-6670\AA\ region,
	using DIB parameters from \citet{JD94}.
	}
\end{figure}

\subsection{Diffuse Interstellar Bands\label{sec:DIBs}}

The observed extinction curve includes a large number of weak extinction
features, known as the ``diffuse interstellar bands'', or DIBs.
The DIBs have FWHM $\gtsim 1\Angstrom$, too broad to be due to molecules
with $\ltsim 5$ atoms in the gas phase, and are therefore features of
the grain population.
\citet{JD94}
list 154 ``certain'' DIBs
in the interval 0.38--0.868$\micron$, plus an additional 52 ``probable''
detections.
Figure \ref{fig:dibs} shows an average extinction curve for
$1.5\micron^{-1} < \lambda^{-1} < 1.75\micron^{-1}$, a wavelength range
which includes several strong DIBs, most notably the DIB at 5780\AA.
This curve consists of the smooth extinction calculated using the \citet{Fi99}
parametrization plus individual diffuse bands
approximated by Drude profiles with central wavelengths, FWHM, and
strengths from the tabulation of Jenniskens \& Desert.

The first 
DIBs were discovered over 80 years ago 
\citep{He22}
and their interstellar
nature was established 69 years ago
\citep{Me34}, 
yet to date not a single DIB has been convincingly identified!
Some of the DIBs may be absorption features
produced by individual PAH molecules in the interstellar PAH mixture.
\citet{DJD95}
report a positive correlation between
the strength of the 2175\AA\ bump and the strength of certain DIBs.
High resolution observations of the 5797\AA\ and 6614\AA\ DIBs show 
fine structure which could be indicative of rotational
bands 
\citep{KHM96,KHF98}.

DIBs have recently been detected toward reddened stars in the LMC and SMC
\citep{ECJ02}.

If the DIBs were due to absorption in the aligned grains responsible for
the polarization of starlight (\S\ref{sec:pol}) there should be
excess polarization associated with the DIBs 
\citep{MA74,PS77}.
All attempts to measure polarization structure associated with DIBs have
yielded only upper limits 
\citep[see][]{AW95} 
that appear to require that at least the DIBs that have been studied 
originate in non-aligned material, consistent with a molecular origin.

At this time the carriers of the DIBs remain unknown, but it seems likely
that at least some are due to large molecules/ultrasmall
grains, possibly of PAH composition.

\subsection{3.4$\mu$m Feature: Aliphatic C-H Stretch\label{sec:CH}}

A broad extinction feature at 3.4$\micron$ is measurable on sightlines
with $A_V\gtsim10~\magn$
\citep{AWD90}.
The feature appears to be present in diffuse atomic regions (in contrast
to the ``ice'' features, see \S\ref{sec:ices}), and therefore is likely
due to ``refractory'' grain material.
The C-H stretching mode in aliphatic (chain-like) hydrocarbons occurs
at 3.4$\micron$.
Since such hydrocarbons are plausible grain constituents, the
observed 3.4$\micron$ absorption is generally attributed to the aliphatic
C-H stretch 
\citep{SAT91}.

Unfortunately, aliphatic hydrocarbons show considerable variation in the
C-H band strength, so it has not been possible to use the measured 
3.4$\micron$ feature to 
determine the abundance of
aliphatic hydrocarbon material in the interstellar
grain population, nor is it possible to identify the specific hydrocarbon
material 
\citep{PA02}.
\citet{SAT91,SPA95}
suggest that the 3.4$\micron$ feature in the
diffuse ISM is due mainly to short saturated aliphatic chains
containing $\sim$4\% of the total C abundance.
\citet{GLM95}
suggest that the 3.4$\micron$ feature is due to carbonaceous organic residues
produced by UV photolysis of ice mantles; 
\citet{DSS98}
attribute the feature to hydrogenated amorphous carbon (HAC) material 
incorporating 20--25\% of the total C abundance;

In grain models where the 3.4$\micron$ absorption 
is assumed to arise in hydrocarbon mantles coating silicate grains
\citep[e.g.,][]{LG97},
the 3.4$\micron$ feature would be
polarized when the 9.7$\micron$ silicate feature is polarized;
this allows a direct test of the
silicate core/hydrocarbon mantle model for interstellar dust
\citep{LG02}.

\subsection{Ice Features\label{sec:ices}}

Sightlines passing through dense molecular clouds typically show additional
absorption features.
The strongest feature is the $3.1\micron$ O-H stretching mode in solid
H$_2$O.
\citet{WBL88}
showed that in the Taurus dark cloud complex, 
the strength of this feature is approximately given by
\begin{equation}
\Delta \tau_{3.1} \approx \left\{ 
\begin{array}{ll}
0	& {\rm for~} A_V \ltsim 3.3~{\rm mag}~~,\\
0.093(A_V-3.3{\rm mag})	& {\rm for~} A_V \gtsim 3.3~{\rm mag}~~,\\
\end{array}
\right.
\end{equation}
suggesting that dust in regions with $A_V>3.3$ has an ice coating,
whereas dust in regions with $A_V<3.3$ is iceless.
The 3.1$\micron$ feature is accompanied by a number of weaker
features due to H$_2$O, plus additional features due to CO$_2$, NH$_3$,
CO, CH$_3$OH, CH$_4$, and other species.
The features due to different species often overlap, and the positions
and shapes
of particular spectral features can be sensitive to the composition of
the frozen mixture, complicating abundance determinations.
Table \ref{tab:ices} gives the abundances relative to H$_2$O estimated
for two sightlines into dense star-forming regions (W33A and W3:IRS5)
and for the sightline to Sgr A$^*$ at the Galactic Center.
Although all three sightlines have similar CO$_2$/H$_2$O ratios, the 
abundances of other species (e.g., CH$_3$OH) vary considerably from
one sightline to another.
The composition of the ice mantle must be sensitive to both local environment
and history.

\newcommand	\hho	{a}
\newcommand	\chhhoh {b}
\newcommand	\coo	{c}
\newcommand	\gibb	{d}
\newcommand	\gerak	{e}
\newcommand	\gurtler{f}
\newcommand	\lacy	{g}
\newcommand	\chiar	{h}
\newcommand	\yyy	{i}
\newcommand	\keane	{j}
\newcommand	\nhhh	{k}
\newcommand	\taban	{l}
\begin{table}[h]
\caption{Major Ice Components}
\label{tab:ices}
{\footnotesize
\begin{center}
\begin{tabular}{@{}c c c c c c c@{}}
\toprule

X	&$\lambda$
		&$\Delta\tau_\lambda$
		&\multicolumn{3}{c}{$N({\rm X})/N({\rm H_2O})$}\\
		\cline{4-6}
	&($\mu$m)	
		&W33A
		&W33A	&W3:IRS5	&GC	\\
\toprule
H$_2$O	
	&3.1$^{\hho}$
		&$5.5\pm0.5$$^{\gibb}$	
		&1
		&1
		&1 \\
XCN	& 4.62
	&1.4$^{\gibb}$
		&.035$^{\gibb}$
		&
		&\\
CO
	&4.67
	&1.3$^{\gerak}$
		&.081$^{\gerak}$	
		&.030$^{\gerak}$
		&$<0.12^{\chiar}$\\
H$_2$CO	&5.81$^{\keane}$
	&0.17$^{\keane}$
		&.065$^{\keane}$
		& -
		&$<.024$$^{\chiar}$\\  
CH$_4$	&7.7		
	&0.11$^{\gurtler}$	
		&.016$^{\gibb}$	
		&.0043$^{\gurtler}$ 
		&.024$^{\chiar}$ \\  
NH$_3$	
	&9.0$^{\nhhh}$	
	&				
		&$<0.05^{\taban}$	
		&$<.035$$^{\gurtler}$	
		&0.2-0.3$^{\lacy}$\\	
CH$_3$OH
	&9.8$^{\chhhoh}$	
	&$0.9\pm0.2$$^{\gibb}$	
		&0.18$^{\gibb}$	
		&$<.004$$^\gurtler$ 
		&$<.04$$^{\chiar}$\\ 
CO$_2$
	&15.2$^{\coo}$
	&0.58$^{\gerak}$	       %
		&0.13$^{\gerak}$	
		&0.13$^{\gerak}$	
		&0.14$^{\gerak}$\\ 
\botrule
\end{tabular}\\
\end{center}
$^{\hho}$ Also 6.0, 13.5, 45$\mu$m\\
$^{\chhhoh}$ Also 4.27$\mu$m\\
$^{\coo}$ Also 3.53, 6.85$\mu$m\\
$^{\gibb}$ \citet{GWS00}\\
$^{\gerak}$ \citet{GWE99}\\
$^{\gurtler}$ \citet{GKH02}\\
$^{\lacy}$ \citet{LFS98}\\
$^{\chiar}$ \cite{CTW00}\\
$^{\keane}$ \cite{KTB01}\\
$^{\nhhh}$ Also 2.27, 2.96, 3.48$\mu$m\\
$^{\taban}$ \cite{TSP03}\\
}
\end{table}

The grains in star-forming clouds are often aligned (see \S\ref{sec:IRpol}),
and the radiation reaching us from embedded sources is often polarized.
Polarization has been measured in the features at 3.1$\micron$ (H$_2$O),
4.6$\micron$ (XCN) and 4.67$\micron$ (CO) 
\citep{CHW96,HCM96}.

\begin{figure}[htb]
\begin{center}
\epsfig{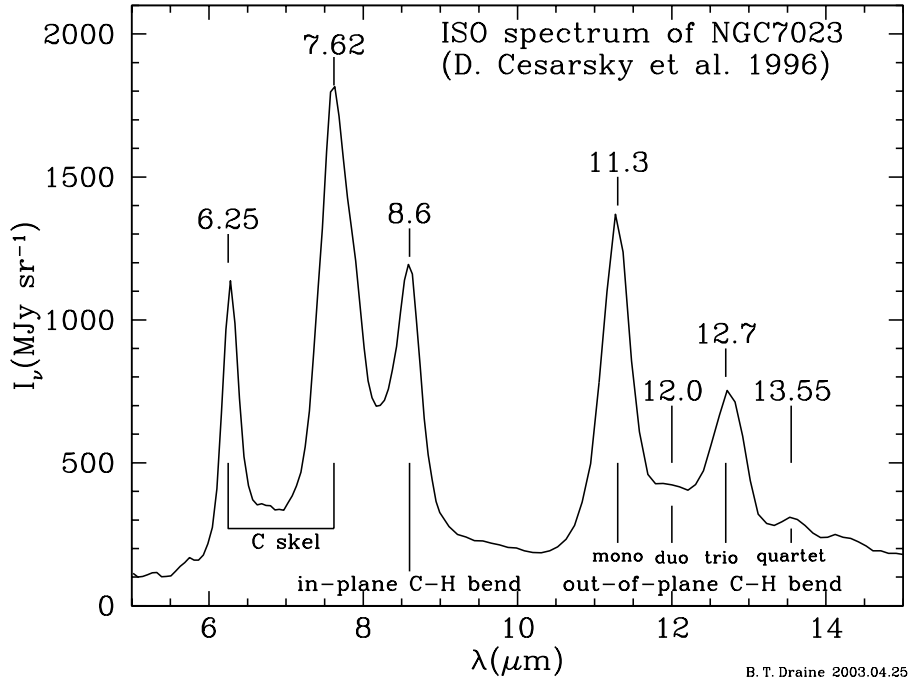}
\vspace*{-0.3cm}
\end{center}
%
%
%
\caption{\footnotesize
	\label{fig:7023}
	PAH emission features in the 5--15$\micron$ 
	spectrum of the reflection nebula NGC 7023.
	}
\end{figure}

\begin{figure}[htb]
\begin{center}
\epsfig{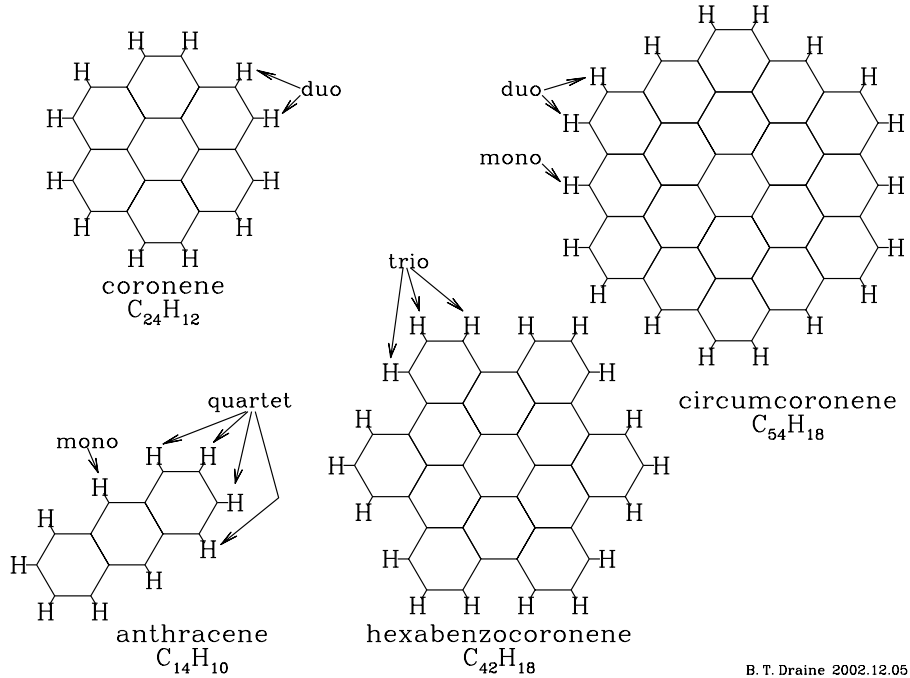}
\vspace*{-0.3cm}
\end{center}
%
%
%
\caption{\footnotesize
	\label{fig:pah}
	The structure of 4 PAH molecules.
	Examples of mono, duo, trio, and quartet H sites are indicated.
	}
\end{figure}

\subsection{3.3, 6.2, 7.7, 8.6, 11.3$\mu$m PAH Features\label{sec:PAHs}}

Many emission nebulae (HII regions, planetary nebulae) and reflection
nebulae
show emission in the 3--15$\micron$ region that is far stronger than
expected from $a\gtsim 100\Angstrom$ grains heated by the ambient
radiation field and emitting thermally.
Much of the emission is concentrated in five
features at 3.3, 6.2, 7.7, 8.6, and 11.3$\micron$
(see Figure \ref{fig:7023}).
\citet{LP84}
identified these features as the optically-active
vibrational modes of PAH molecules.
When H atoms are attached to the edge of an aromatic ring skeleton,
there are characteristic optically-active vibrational modes
\citep{ATB89}:
\begin{itemize}
\item C-H stretching mode at 3.3$\micron$
\item C-C stretching mode at 6.2$\micron$
\item C-C stretching mode at 7.7$\micron$
\item C-H in-plane bending mode at 8.6$\micron$
\item C-H out-of-plane bending mode with wavelength
	depending on the number of neighboring H atoms:
	\begin{itemize}
	\item 11.3$\micron$ for ``mono'' H (no adjacent H)
	\item 12.0$\micron$ for ``duo'' H (2 contiguous H)
	\item 12.7$\micron$ for ``trio'' H (3 contiguous H)
	\item 13.55$\micron$ for ``quartet'' H (4 contiguous H)
	\end{itemize}
\end{itemize}
Examples of
``mono'', ``duo'', 
``trio'', and ``quartet'' sites are indicated in Figure \ref{fig:pah}. 
Although the broadness of the observed features precludes identification
of specific PAH molecules, a population of PAH molecules appears to
provide a natural explanation for the observed emission spectrum,
where the vibrational excitation is assumed to be the result
of ``internal conversion'' of energy following absorption of
an optical or UV photon
\citep{LP84,ATB89}.

There should in principle be absorption associated with these emission 
features.
An absorption feature at 6.2$\micron$ (see Figure \ref{fig:ext_ir})
has been observed 
\citep{SHW98,CTW00}.
It matches closely the
expected $6.2\micron$ absorption feature 
\citep{LD01b}.

\subsection{X-Ray Absorption Edges\label{sec:xrayedge}}

Interstellar grains absorb and scatter X-rays.
The X-ray absorption by atoms such as C, O, Mg, Si, and Fe will show
photoelectric ``absorption edges'' similar to atoms in the gas phase,
but the precise energy and structure of the absorption edge will be dependent
on the chemical nature of the grain material
\citep{FWC98,Dr03b}.
Significant structure in both scattering and absorption is expected
near the C~K (284~eV), O~K (538~eV),
Fe~L$_{2,3}$ (708, 721~eV), Mg~K (1311~eV), 
Si~K (1846~eV), and Fe~K (7123~eV) edges \citep{Dr03b}.
The Chandra observatory has been used to study the absorption edges of O 
\citep{PBM01,TFM02}
and O, Mg, Si, and Fe 
\citep{SCC02}
but it has not yet proved possible to identify the chemical form in which
the solid-phase Mg, Si, Fe and O are bound.

\subsection{Extended Red Emission\label{sec:ERE}}

The ``extended red emission'' (ERE) 
from dust grains provides
a potentially important clue to the composition of 
interstellar dust.
The ERE consists of a broad featureless emission band between
$\sim5400\Angstrom$ and $\sim9000\Angstrom$, peaking at
$6100\Angstrom\ltsim\lambda_p\ltsim8200\Angstrom$, and with a FWHM in the
range 600--1000\AA.
The peak wavelength $\lambda_p$ and band profile appear to vary
from one region to another.

First observed in the Red Rectangle 
\citep{SCM80},
ERE has since been seen in a wide variety of dusty
environments, including reflection nebulae 
\citep{WS85,WB90},
planetary nebulae 
\citep{FW90},
HII regions
\citep{DZP00},
the diffuse ISM of our Galaxy 
\citep{GWF98,SG98},
and from other galaxies 
\citep{PMB02}.

The ERE must be
photoluminescence: absorption of a starlight photon followed
by emission of a lower energy photoluminescence photon.
Based on the detection of ERE from the diffuse ISM,
the photon conversion efficiency of interstellar dust
is estimated to be $10\%\pm3\%$ 
\citep{GWF98,SG98}.
The photoluminescence efficiency of the ERE carrier must exceed this, since
it is presumably not the only UV/visible photon absorber.

The luminescing substance remains uncertain.
Candidate ERE materials include 
HAC 
\citep{Du85,WS88},
PAHs 
\citep{dLO86},
quenched carbonaceous composite 
\citep{SWN92},
C$_{60}$
\citep{We93}, 
coal 
\citep{PCG96},
silicon nanoparticles 
\citep{LEG98,WGF98,SW02},
and carbon nanoparticles 
\citep{SD99}.

Most candidate materials appear unable to match the observed ERE spectra and
efficiency 
\citep{WGF98}.
Lab studies of the luminescence spectrum of hydrogenated amorphous carbon (HAC)
appeared to show a good match to observed ERE spectra 
\citep{WS88}
provided the HAC is annealed sufficiently to shift the luminescence into the
red, where it is observed; however, such red-luminescing HAC has a very low
photoluminescence efficiency 
\citep{FW93,RRA96}
that is incompatible with the high efficiency required to explain the
ERE from the diffuse ISM.

Each of the nanoparticle hypotheses appears to have difficulties:
(1) 
\citet{SD99}
argue that small carbon particles with
mixed $sp^2/sp^3$ bonding can reproduce the
observed ERE profile and required efficiency, but they
predict a second ERE peak at $\sim1\micron$ which does not appear
to be present in NGC 7023 
\citep{GWR00}.
(2) The PAH hypothesis has difficulty with 
(a) ERE emission and PAH emission having differing
spatial distributions in HII regions 
\citep{SP93,DZP00},
and (b) nondetection of ERE in reflection nebulae illuminated by stars
with $T_{\rm eff}<7000\K$ 
\citep{DPS99},
whereas PAH emission bands are seen in these regions.
(3) The silicon nanoparticle hypothesis appears to be ruled out by
nondetection of $\sim20\micron$ emission that should result from single-photon
heating (see \S\ref{sec:IRemission})
in the silicon nanoparticle model 
\citep{LD02a}.

Recently, two new emission features have been reported in the near-infrared,
at $1.15$ and $1.5\micron$ 
\citep{GWR00}.
\citet{SGC01}
propose that the $1.5\micron$ feature is due to $\beta$-FeSi$_2$.

Although it would appear that all current proposals have been ruled out,
the arguments against them are not yet conclusive.
For example, perhaps it is possible to prepare HAC samples that
achieve a higher photoluminescence efficiency, or there may be
some way to explain the lack of PAH emission seen in the Bubble nebula,
or silicon nanoparticles might be part of larger structures.
At this point it seems appropriate to continue consideration of
HAC, PAHs, and silicon nanoparticles as candidate ERE carriers.

\section{PRESOLAR GRAINS IN METEORITES AND INTERPLANETARY DUST
\label{sec:meteor}}

The most primitive meteorites contain presolar grains, which are recognized
by virtue of isotopically anomalous composition -- see 
\citet{HZ00}
for a recent review.
Table \ref{tab:meteorites} lists the 
six presolar grain types that have been identified thus far, along with
their abundance (fraction of the total mass) in the bulk meteorite.
Nanodiamond 
($sp^3$-bonded carbon) grains make up 0.05\% = 500~ppm of the meteorite.
A tiny fraction of the nanodiamond particles contain a trapped Xe atom;
the isotopic abundances of the trapped Xe suggests formation in SN ejecta.
SiC grains are the next most abundant by mass (6~ppm), 
with isotopic patterns that
indicate formation primarily in outflows from AGB stars.
About 0.5\% of the SiC grains are of ``Type X'', 
with isotopic abundances, including live
$^{44}$Ti ($T_{1/2}=$~59~yr) at the time of grain formation, 
which suggest formation in Type II supernovae.
About 1 ppm of the meteorite consists of graphitic 
($sp^2$-bonded carbon) grains, formed in AGB stars, supernovae, and novae.
Al$_2$O$_3$ corundum grains are found, at least some of which appear
to have formed in outflows from red giants and AGB stars.
Finally, there are also Si$_3$N$_4$ grains which appear to have a SN II
origin.

\newcommand	\qqq	{a}
\newcommand	\ooo	{b}
\newcommand	\ppp	{c}
\newcommand	\rrr	{d}

\begin{table}[ht]
\caption{Presolar Grains in Meteorites \citep{HZ00}}
\label{tab:meteorites}
{\footnotesize
\begin{center}
\begin{tabular}{
	@{} l 
	@{\hspace*{1.0em}} c 
	@{\hspace*{1.0em}} c 
	@{\hspace*{1.0em}} c
	@{}}
\toprule
Composition	& diameter($\micron$)	&Abundance$^{\qqq}$ &Origins$^{\ooo}$\\
\toprule
C(diamond)	& 0.002	&$5\times10^{-4}$		&SN\\
SiC$^{\ppp}$		& 0.3--20&$6\times10^{-6}$	&AGB\\
C(graphite)$^{\rrr}$	& 1--20&$1\times10^{-6}$		&AGB,SN~II,nova\\
SiC type X	& 1--5&$6\times10^{-8}$			&SN\\
Al$_2$O$_3$ (corundum)	&0.5-3	&$3\times10^{-8}$	&RG,AGB\\
Si$_3$N$_4$	&$\sim$1	&$2\times10^{-9}$	&SN~II\\
\toprule
\end{tabular}

$^{\qqq}$ Overall abundance in primitive carbonaceous chondrite meteorites.\\
$^{\ooo}$ SN = supernova; AGB = asymptotic giant branch star;
	RG = red giant\\
$^{\ppp}$ SiC grains sometimes contain very small TiC inclusions.\\
$^{\rrr}$ graphite grains sometimes contain very small
TiC, ZrC, and MoC inclusions.\\
\end{center}
}
\end{table}

It is important to recognize that the procedures used to search for
presolar grains in meteorites
begin with chemical treatments designed to dissolve the silicate matrix in
which the presolar grains are embedded.
It therefore comes as no surprise that silicates are not present in 
Table \ref{tab:meteorites}.
It should also be recognized that the formation of even the most primitive
meteorites involved temperatures high enough to melt rocks, in an
environment which was presumably oxidizing (since the solar nebula 
had O/C $>$ 1).
Therefore the relative abundances in Table \ref{tab:meteorites}
probably have more to do
with the ability of grain types to survive a period of exposure to a hot,
oxidizing environment than with interstellar abundance.

Much more informative would be the abundances of presolar grains in
cometary material.  A ``core sample'' would be ideal, but at this time there is
no planned mission for returning a sample of pristine cometary ice to Earth.
However, the Stardust mission 
\citep{BTC00}
will collect cometary dust on an aerogel panel 
during the $6.1~\kms$ encounter of the
Stardust spacecraft with Comet Wild 2 in January 2004, and return it to Earth.

Interstellar grains enter the solar system, and have been detected by
the Ulysses and Galileo spacecraft; the impact rate is consistent
with expectations for $a\approx0.3\micron$ grains, but the impact
rate for $a\gtsim 1\micron$ grains is much larger than expected for
the average interstellar dust distribution 
\citep{FDG99,WD01a}.
The Stardust mission will expose an aerogel panel to
interstellar grains flowing through the solar system; with $\sim 25\kms$
impact speeds, these grains are likely to be destroyed, but their residue in
the aerogel can be analyzed for total mass and composition.
We await the return of Stardust in January 2006.

Active comets release their dust grains, which become part of the 
interplanetary dust particle (IDP) population.
IDPs can be collected from
the Earth's atmosphere 
\citep{Br85}.
\citet{Br94}
identifies a class of glassy silicate 
IDPs, known as GEMS (Glass with Embedded Metals and Sulfides),
which appear to be presolar.
Most GEMS are between 0.1 and 0.5$\micron$ in diameter -- similar to the
sizes of interstellar silicates (see Figure \ref{fig:dnda}).
The $8-13\micron$ infrared absorption spectrum of these grains is similar
to interstellar silicates 
\citep{BKS99}.

\section{CANDIDATE GRAIN MATERIALS}

Based on the observations and clues described above, it is possible to
formulate a short list of candidate materials for the bulk of interstellar
dust.  We will omit from consideration ``rare'' materials like Al$_2$O$_3$ --
even though Al is depleted from the gas phase, 
given the solar abundance of Al (Al/H $= 3.1\times10^{-6}$) 
aluminum-based materials account for only a few \% of interstellar
grain mass.

\subsection{Silicates\label{sec:cand_silicate}}

There is little doubt that silicate material contributes a substantial
fraction of the total mass of interstellar dust.
As discussed above (\S\ref{sec:silicate}), in the ISM,
at least $\sim$95\% of the silicate material is amorphous
\citep{LD02a}.
In some circumstellar disks the crystalline fraction appears to be higher,
although it appears that the bulk of the silicates remain amorphous.
\citet{BMK01}
find that 5-10\% of the olivine mass is crystalline
in the dust around Herbig Ae/Be stars.

Cosmic abundances dictate that the dominant metal ions in silicates will be
either Mg, Fe, or both.  There are two general chemical classes of crystalline
silicates:
Mg-Fe pyroxenes Mg$_x$Fe$_{1-x}$SiO$_3$ (including enstatite MgSiO$_3$ and
ferrosilite FeSiO$_3$)
and olivines Mg$_{2x}$Fe$_{2-2x}$SiO$_4$ (including forsterite Mg$_2$SiO$_4$
and fayalite Fe$_2$SiO$_4$).
There are two different crystalline structures for the pyroxenes: 
orthopyroxene and clinopyroxene.

The optical constants of various
crystalline pyroxenes have been measured by
\citet{JMD98}
and
\citet{CKT02};
crystalline olivines have been measured by
\citet{JMD98},
\citet{KTS99},
and
\citet{SKS02}.

As noted above, some circumstellar dust exhibits emission features 
characteristic of crystalline silicates,
and from the locations and strengths of these
features it is possible
to infer the Mg:Fe ratio.
In all cases thus far the crystalline silicate material
appears to be very Mg-rich and Fe-poor,
consistent with pure forsterite and enstatite
\citep{TWM98,MWT02a}.
\citet{MWT02b}
report that enstatite MgSiO$_3$ is more abundant
than forsterite Mg$_2$SiO$_4$ around most evolved stars.

Si is generally heavily depleted in the ISM. 
\underbar{If} Si is predominantly in
silicates, and \underbar{if} Fe is not present in silicates, 
then the interstellar
silicates would be expected to be enstatite MgSiO$_3$.  The Fe would then
presumably be in either metallic Fe or in oxide form 
(FeO, Fe$_2$O$_3$, Fe$_3$O$_4$).
However, while there is evidence for Mg-rich silicate in circumstellar
dust, we do not yet have a reliable indication of the
composition of {\it interstellar} silicates, so an appreciable
Fe fraction, and olivine-like composition, are not ruled out.

Observed gas-phase abundances of Mg, Fe, and
Si, and variations in these abundances from one cloud to another,
have been interpreted as indicating that silicate grains may have a
Mg-rich mantle and Fe-rich core 
\citep{SF93}.
The total ratio of (Fe+Mg) atoms to Si atoms in grains is estimated
to be $\sim$3.4:1 
\citep{Fi97},
$\sim$3:1 
\citep{HSF99} -- in
excess of the 2:1 ratio expected for olivines, 
suggesting that metal oxides or
metallic Fe may be present in addition to silicates.
However, these conclusions depend upon assumed values of the total
interstellar abundances; 
\citet{SM01}
conclude that the (Fe+Mg):Si ratio in interstellar dust is
actually close to 2:1 for both ``halo'' dust and the dust in the
well-studied $\zeta$Oph cloud.

The composition of interstellar silicates remains uncertain,
but the study by 
\citet{SM01}
suggest that olivine MgFeSiO$_4$ appears to be a reasonable approximation.

\subsection{Carbonaceous Materials}

By ``carbonaceous'' we mean materials which are predominantly C by mass,
including:
\begin{itemize}
\item Diamond: $sp^3$-bonded carbon.
\item Graphite: $sp^2$-bonded carbon, either monocrystalline 
	or polycrystalline.
\item Amorphous or glassy carbon: a mixture of $sp^2$- and $sp^3$-bonded 
	carbon, with only short-range order.
\item Hydrogenated amorphous carbon: amorphous carbon with an appreciable
	hydrogen fraction.
\item Polycyclic aromatic hydrocarbons: $sp^2$-bonded carbon with 
	peripheral H (see \S\ref{sec:PAHs}).
\item Aliphatic (chainlike) hydrocarbons (see \S\ref{sec:CH})
\end{itemize}
The presence of $sp^2$-bonded carbon is indicated by the 2175\AA\ feature
and the PAH emission bands.
The 2175\AA\ feature requires $sp^2$-bonded carbon in small
($a\ltsim100\Angstrom$) particles (in larger particles the 2175\AA\ feature
is suppressed), and single-photon excitation of the 
PAH emission features requires very small particles, with $\ltsim10^4$ C
atoms.
A PAH population incorporating a C abundance
C/H$\approx 60\times10^{-6}$ appears able to account for both 
the PAH emission features (\S\ref{sec:PAHs})
and the 2175\AA\ feature (\S\ref{sec:2175}).

\subsection{SiC}

SiC grains are found in meteorites (\S\ref{sec:meteor}) 
and an emission feature at 11.3$\micron$ is observed in the 
spectra of many carbon stars
\citep{TC74,BBF98}.
However, absence of the 11.3$\micron$ feature in the interstellar extinction
indicates that the abundance of Si in SiC dust is less than 5\% 
of the abundance of Si in silicates 
\citep{WDM90}.
Accordingly, SiC grains, although undoubtedly present in the ISM,
are not a major component of the interstellar grain mix.

\subsection{Carbonates}

Calcite CaCO$_3$ and dolomite CaMg(CO$_3$)$_2$ 
have been detected in dusty disks within
the planetary nebulae NGC6302 and NGC6537 
\citep{KJW02},
but these carbonates are estimated to contribute less than 1\% of the
dust mass.
Carbonates are evidently not a major component of the interstellar
grain mix.

\section{POLARIZATION
	\label{sec:pol}}

\subsection{Optical-UV Polarization\label{sec:pol_optical}}

Light reaching us from reddened stars is often linearly polarized, because the
extinction depends on the polarization mode.
The degree of polarization as a function of wavelength
can be approximated by the
``Serkowski law'', a 3-parameter empirical fitting function 
\citep{Se73}:
\beq
p(\lambda) = p_{\rm max}\exp\left[-K (\ln(\lambda/\lambda_{\rm max}))^2\right]
~~~,
\eeq
where $\lambda_{\rm max} \approx 5500\Angstrom$ and $K\approx 1$
on typical sightlines through diffuse clouds.
The polarization arises from differential extinction by aligned dust
grains, and will depend on the degree of
alignment with the local magnetic field, the angle between the local
magnetic field and the line of sight, and the degree to which the
direction of the magnetic field varies along the line of sight.
Studies of many sightlines 
\citep{SMF75}
find that 
\beq
p_{\rm max} \ltsim 0.03 A(\lambda_{\rm max})/\magn
~~~,
\eeq
where $A(\lambda)$ is the (polarization-averaged) extinction at
wavelength $\lambda$.
Interstellar dust grains must be sufficiently
nonspherical and sufficiently aligned so that at $\lambda_{\rm max}$
one polarization mode
is extincted $\sim$6\% more than the other polarization mode.
Values of $p_{\rm max}/A(\lambda_{\rm max}) < 0.03/\magn$ are presumed to
result when the magnetic field is disordered or not transverse to
the line-of-sight, or from regions where the degree of grain alignment
is lower.

The wavelength $\lambda_{\rm max}$ varies from one sightline to another,
and is correlated with the value of $R_V$, with
\beq
R_V \approx 3.67(\lambda_{\rm max}/5500\Angstrom) - 0.29
\eeq
\citep{CM88},
and
\beq
K \approx 1.02 (\lambda_{\rm max}/5500\Angstrom) - 0.10
\eeq
\citep{WLR82}.
The Serkowski law provides quite a good fit to the data for 
$0.8\micron^{-1} \ltsim \lambda \ltsim 3\micron^{-1}$.
Observations in the vacuum UV 
\citep{AWC96,CWA96,WCK97}
show that the degree of polarization continues to decline
with decreasing wavelength.
This implies
that the small grains responsible for the rising extinction in
the vacuum UV are inefficient
polarizers -- they are either nearly spherical or minimally aligned
\citep{KM95}.

Most stars show no polarization excess in
the 2175\AA\ feature but two do: a
$2175\Angstrom$ polarization feature is detected in HD~197770 and HD~147933-4
\citep{WCK97}.

\subsection{Polarized Far-Infrared Emission \label{sec:IRpol}}

The aligned dust grains that polarize starlight will also produce
linearly-polarized infrared emission, because for an isotropic dielectric
function a submicron grain will radiate most effectively when the
E field is parallel to the ``long'' axis of the grain.
It is now possible to map this polarized emission from dense regions.
For example, the W51 star-forming region has been mapped at 100$\micron$
\citep{DDD00},
850$\micron$ 
\citep{CAJ02},
and 1.3mm 
\citep{LCG01},
with linear polarizations as large as 
10\% observed at 1.3mm.

\citet{HDD99}
have discussed the wavelength-dependence of the
measured polarization for dense cloud cores and envelopes.  
The data suggest that the
grain alignment is larger in warmer regions.

Observations of quiescent clouds are consistent with no grain alignment
in the central regions 
\citep{GJL95,PGD01}.

Radiative torques due to anisotropic starlight play an important role
in the process of grain alignment 
\citep{DW97}.
These torques would be ineffective in the central regions of quiescent
dark clouds where external starlight is heavily attenuated.
In star-forming clouds such as M17 or W51, however, starlight is provided
within the clouds by recently-formed stars.

\section{A PROVISIONAL GRAIN MODEL
	\label{sec:prov_model}}

Although we have quite good observational determinations of the extinction
in the near-infrared, optical, and UV regions, the extinction
and scattering properties of interstellar grains at very short wavelengths
and very long wavelengths are probably best obtained by calculating them
using a physical grain model that has been constrained by a variety
of observations, including observations of extinction and infrared emission.

\citet{MRN77}
discovered that the average
interstellar extinction could be satisfactorily reproduced by a grain
model containing two components -- graphite grains and silicate grains.
Remarkably, the extinction curve was reproduced very well if both
grain components had power-law size distributions, $dn/da \propto a^{-3.5}$,
truncated at a minimum size $a_-\approx 50\Angstrom$ and a maximum
size $a_+\approx 2500\Angstrom$.
The success of this grain model led
\citet{DL84}
to refine the optical constants, and extend the
treatment into the mid- and far-infrared.
The graphite-silicate model was applied in the X-ray region
by \citet{LD93}.

At the time when the ``graphite-silicate'' model was developed, 
PAHs had not been
recognized as a major interstellar grain material.
With the recognition that the interstellar grain population appears to
include a substantial population of ultrasmall grains with PAH composition,
the ``graphite-silicate'' model has recently been extended in a very natural
way: the carbonaceous grains are now assumed to be PAH molecules when very
small, but to have physical and chemical properties that can be approximated
by grains of bulk graphite when larger than $\sim 0.01\micron$ (containing
$N \gtsim 10^6 $ C atoms).
The properties of the carbonaceous grains are taken to change smoothly from
PAH-like to graphite-like as the grain size is increased.

Models developed to account for the infrared emission spectrum of interstellar
dust require that the PAH population contain $\sim15\%$ of the interstellar
C abundance.
These particles necessarily make a major contribution to the UV
extinction.
\citet{WD01a}
have found size distributions for the
grains that, including the PAHs, are consistent with the
extinction produced by the interstellar grain population,
and with the size distribution of the PAHs adjusted to reproduce the
observed infrared emission from the diffuse ISM
\citep{LD01b}.

\begin{figure}[htb]
\begin{center}
\epsfig{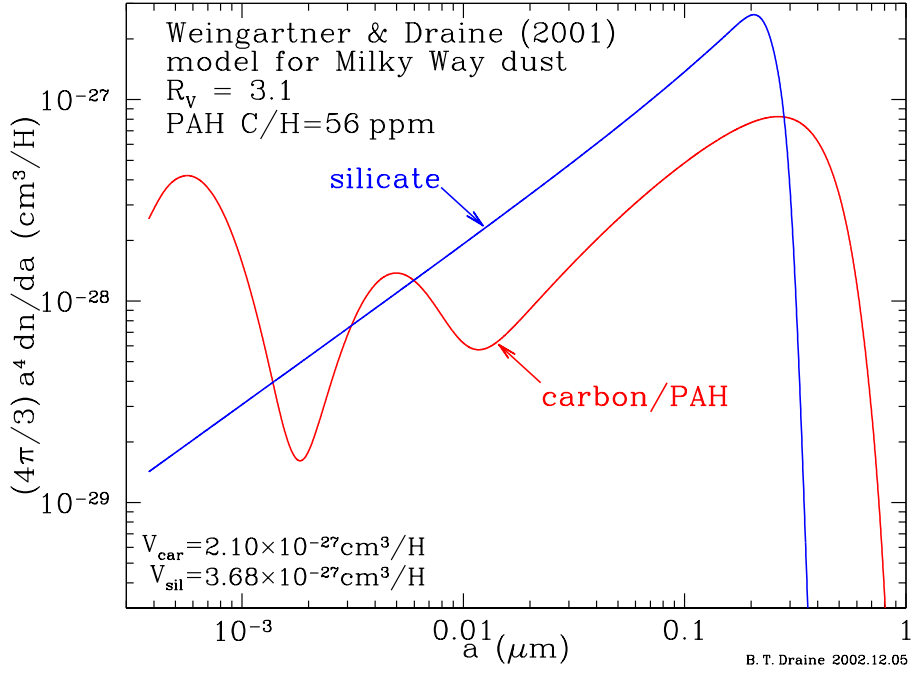}
\vspace*{-0.3cm}
\end{center}
%
%
%
\caption{\footnotesize
	\label{fig:dnda}
	Size distributions 
	for carbonaceous-silicate grain model of
	\citet{WD01a}
	for Milky Way dust with $R_V=3.1$, but
	with abundances decreased by a factor 0.93 (see text).
	}
\end{figure}

The grain size distribution adopted by \citet{WD01a} for average Milky Way
dust (with $R_V=3.1$) is shown in Figure \ref{fig:dnda}, but with
abundances reduced by a factor 0.93 to reproduce $A_{I_C}/N_{\rm H}$ from
eq.\ (\ref{eq:AI_per_H}).
The carbonaceous grain distribution is trimodal:
The peak at $a\approx0.3\micron$ is required to reproduce the observed
extinction curve;
the peak at $a\approx.0005\micron$
is required to reproduce the $3-12\micron$ PAH emission features; and the
peak at $a\approx0.005\micron$ improves the fit to the observed emission near
$60\micron$.
The peaks at $\sim0.3\micron$ and $\sim.0005\micron$ 
are certainly real, but the peak at $.005\micron$
could be an artifact of errors in the adopted grain
optical and FIR cross sections.

\begin{figure}[htb]
\begin{center}
\epsfig{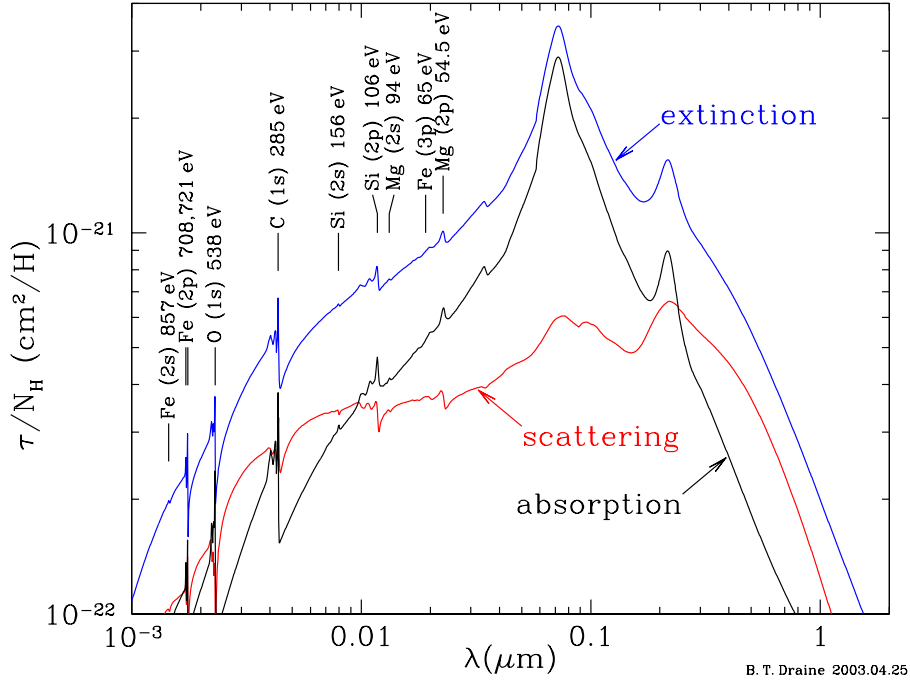}
\vspace*{-0.3cm}
\end{center}
%
%
%
\caption{\footnotesize
	\label{fig:ext_WD01_opt}
	Extinction and scattering
	calculated for \citet{WD01a} model for $R_V=3.1$ Milky Way dust,
	but with abundances reduced by factor 0.93 (see text).
	}
\end{figure}

\begin{figure}[htb]
\begin{center}
\epsfig{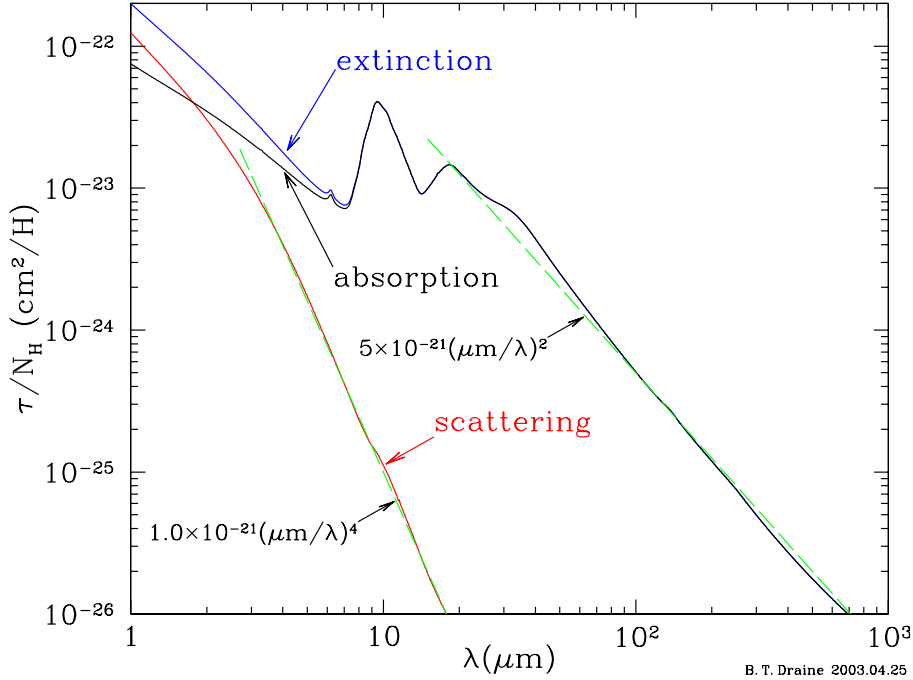}
\vspace*{-0.3cm}
\end{center}
%
%
%
\caption{\footnotesize
	\label{fig:ext_WD01_fir}
	Extinction and scattering 
	calculated for \citet{WD01a} model for $R_V=3.1$ Milky Way dust,
	but with abundances reduced by factor 0.93 (see text).
	The dashed lines show the asymptotic behavior of the
	absorption ($\propto\lambda^{-2}$) and scattering
	($\propto\lambda^{-4}$) cross sections.
	}
\end{figure}

The dielectric functions of graphite and ``astrosilicate''
(including structure at X-ray absorption edges) have been reestimated
by \citet{Dr03b}.
The extinction and scattering cross sections per H nucleon,
calculated for the \citet{WD01a} grain model,
are shown in Figs.\ \ref{fig:ext_WD01_opt} and \ref{fig:ext_WD01_fir}.
For $\lambda\gtsim 20\micron$, 
the extinction calculated for this grain model
can be approximated by 
$\tau/N_{\rm H}\approx 5\times10^{-25}\cm^2(\lambda/100\micron)^{-2}$, as seen
in Figure \ref{fig:ext_WD01_fir}.
The scattering varies as expected for Rayleigh scattering for
$\lambda\gtsim 4\micron$: 
$\tau_{\rm sca}/N_{\rm H}\approx 1.0\times10^{-21}\cm^2 
(\lambda/\micron)^{-4}$.

This provisional dust model does not include a component to 
reproduce the observed 3.4$\micron$
absorption feature (\S\ref{sec:CH}).  
If the graphite in the 
$a\gtsim 200\Angstrom$ carbonaceous grains is replaced with a mixture of
graphite and aliphatic hydrocarbons,
it seems likely that
the extinction curve, including the 3.4$\micron$ feature, 
could be reproduced with only slight adjustments to
the grain size distribution.  

The \citet{WD01a} model has $A_V/\Delta\tau_{9.7}=14.2$ 
(see Tables \ref{tab:model_IRoptuv} -- \ref{tab:model_euvxray}),
25\% smaller than the observed value $18.5$ for diffuse cloud dust
(see Table \ref{tab:AV/tausil}).
Evidently the silicate abundance or the silicate band strength should
be reduced by $\sim 30\%$.

\begin{figure}[htb]
\begin{center}
\epsfig{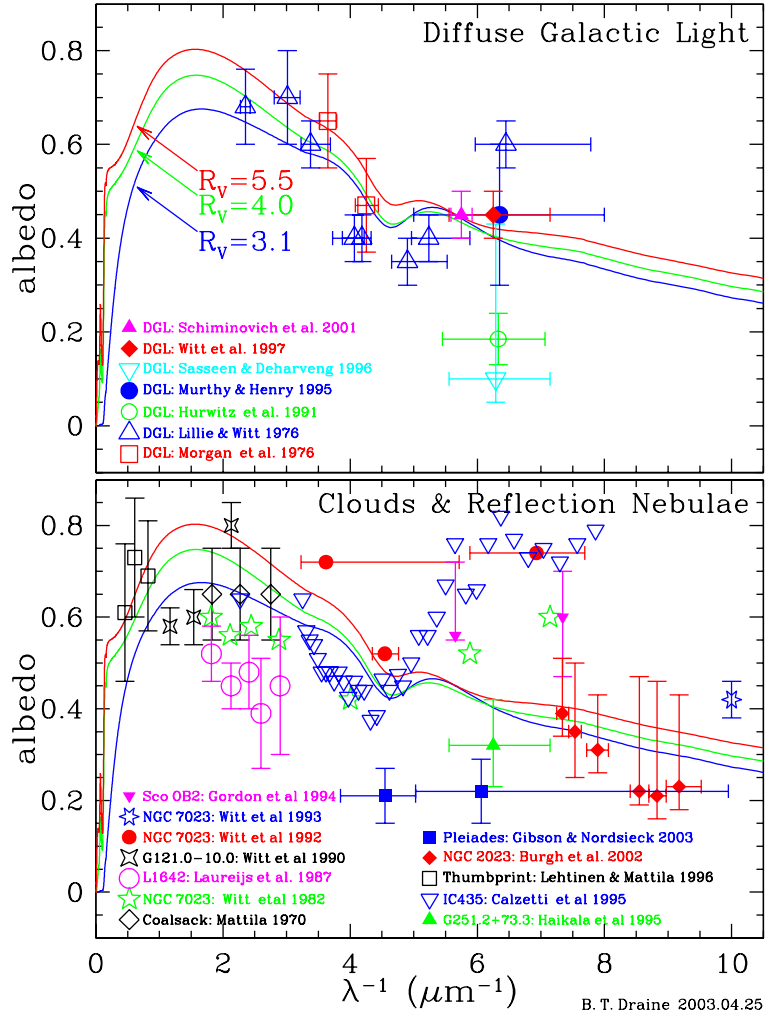}
\vspace*{-0.3cm}
\end{center}
%
%
%
\caption{\footnotesize
	\label{fig:albedo}
	Solid lines: scattering albedo calculated for \citet{WD01a} model for
	Milky Way dust with $R_V=3.1$, 4.0, and 5.5.
	Symbols: observational determinations.
	After \citet{Dr03b}.
	}
\end{figure}

\begin{figure}[htb]
\begin{center}
\epsfig{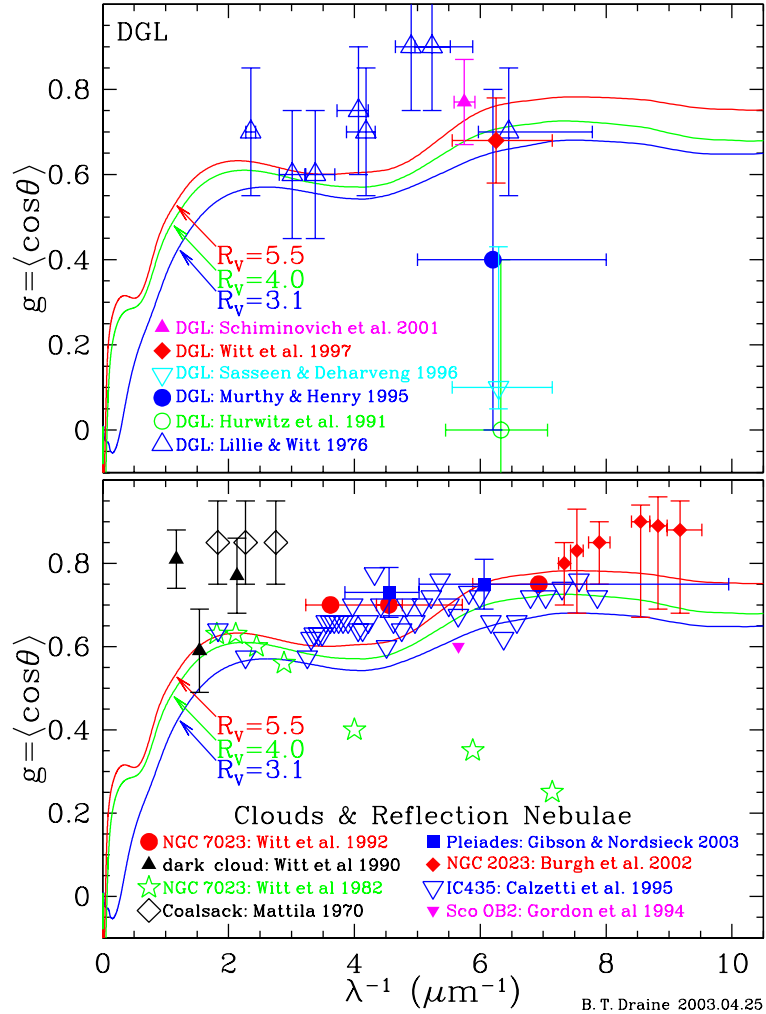}
\vspace*{-0.3cm}
\end{center}
%
%
%
\caption{\footnotesize
	\label{fig:g}
	Solid lines: 
	Scattering asymmetry factor $g=\langle\cos\theta\rangle$
	calculated for \citet{WD01a} model for Milky Way dust 
	with $R_V=3.1$, 4.0, and 5.5.
	Symbols: observational determinations.
	After \citet{Dr03b}.
	}
\end{figure}


\newcommand	\www	{{a}}
\newcommand	\ccm	{{b}}
\newcommand	\fmod	{{c}}
\newcommand	\mmm	{{d}}
\newcommand	\aaa	{{e}}
\newcommand	\bbb	{{f}}
\newcommand	\ccc	{{g}}
\begin{table}
\def~{\hphantom{0}}
\caption{Absorption and Scattering for $5\micron>\lambda>0.1\micron$ $^\www$}
\label{tab:model_IRoptuv}
{\footnotesize
\begin{center}
\begin{tabular}{
	@{}c
	@{\hspace*{0.7em}}c
	@{\hspace*{0.7em}}c
	@{\hspace*{0.7em}}c
	@{\hspace*{0.7em}}c
	@{\hspace*{0.7em}}c
	@{\hspace*{0.7em}}c
	@{\hspace*{1.0em}}c
	@{\hspace*{1.0em}}c
	@{}}
\toprule

\colrule

&&CCM89$^\ccm$&F99$^\fmod$&\multicolumn{5}{c}{model$^\mmm$}\\
\cline{5-9}
$\lambda$&band$^\aaa$&$A_\lambda/A_{I_C}$&$A_\lambda/A_{I_C}$&$A_\lambda/A_{I_C}$&albedo&$\langle\cos\theta\rangle$&$C_{\rm ext}^{\ \bbb}$&$\kappa_{\rm abs}$\\
$\mu$m&&&&&&&cm$^2$/H&cm$^2$/g\\
\toprule
 4.750&$M$&&.0619 &.0456& 0.164&-.041&$1.29\!\times\!10^{-23}$&$5.76\!\times\!10^{  2}$\\
 3.800&$L^{\prime}$&&.0866&.0700& 0.245&-.013&$1.97\!\times\!10^{-23}$&$7.97\!\times\!10^{  2}$\\
 3.450&$L$&.0931&0.103 &.0845& 0.282& .006&$2.38\!\times\!10^{-23}$&$9.16\!\times\!10^{  2}$\\
 2.190&$K$&0.190&0.212 &0.197& 0.439& 0.131&$5.56\!\times\!10^{-23}$&$1.67\!\times\!10^{  3}$\\
 1.630&$H$&0.304&0.321 & 0.324& 0.520& 0.209&$9.15\!\times\!10^{-23}$&$2.35\!\times\!10^{  3}$\\
 1.220&$J$&0.485&0.489 & 0.514& 0.585& 0.289&$1.45\!\times\!10^{-22}$&$3.22\!\times\!10^{  3}$\\
 1.000&&0.682& 0.668&0.707& 0.623& 0.363&$1.99\!\times\!10^{-22}$&$4.02\!\times\!10^{  3}$\\
0.9000&&0.792& 0.819& 0.836& 0.642& 0.401&$2.36\!\times\!10^{-22}$&$4.53\!\times\!10^{  3}$\\
0.8930&$z$&0.803&0.830 & 0.847& 0.643& 0.404&$2.39\!\times\!10^{-22}$&$4.57\!\times\!10^{  3}$\\
0.8655&$I_J$&0.855&0.879 &0.889& 0.647& 0.414&$2.51\!\times\!10^{-22}$&$4.73\!\times\!10^{  3}$\\
0.8020&$I_C$&1.000&1.000 & 1.000& 0.658& 0.438&$2.82\!\times\!10^{-22}$&$5.17\!\times\!10^{  3}$\\
0.7480&$i$&1.134& 1.125& 1.111& 0.665& 0.459&$3.13\!\times\!10^{-22}$&$5.62\!\times\!10^{  3}$\\
0.7000&&1.246&1.255 & 1.226& 0.671& 0.479&$3.46\!\times\!10^{-22}$&$6.09\!\times\!10^{  3}$\\
0.6492&$R_C$&1.365&1.419 & 1.368& 0.674& 0.500&$3.86\!\times\!10^{-22}$&$6.72\!\times\!10^{  3}$\\
0.6415&$R_J$&1.381&1.442 & 1.391& 0.675& 0.503&$3.92\!\times\!10^{-22}$&$6.83\!\times\!10^{  3}$\\
0.6165&$r$&1.443& 1.531& 1.472& 0.676& 0.513&$4.15\!\times\!10^{-22}$&$7.21\!\times\!10^{  3}$\\
0.5470&$V$&1.665& 1.805& 1.735& 0.674& 0.538&$4.89\!\times\!10^{-22}$&$8.55\!\times\!10^{  3}$\\ 
0.4685&$g$&2.045&2.238 & 2.114& 0.661& 0.560&$5.96\!\times\!10^{-22}$&$1.08\!\times\!10^{  4}$\\
0.4405&$B$&2.187&2.396 & 2.273& 0.653& 0.565&$6.41\!\times\!10^{-22}$&$1.19\!\times\!10^{  4}$\\
0.3635&$U$&2.550&2.813 & 2.847& 0.616& 0.569&$8.03\!\times\!10^{-22}$&$1.65\!\times\!10^{  4}$\\ 
0.3550&$u$&2.612&2.866 & 2.847& 0.616& 0.569&$8.03\!\times\!10^{-22}$&$1.65\!\times\!10^{  4}$\\
0.3000&&3.005&3.306 & 3.328& 0.582& 0.556&$9.38\!\times\!10^{-22}$&$2.10\!\times\!10^{  4}$\\ 
0.2500&&3.826&4.163 & 4.150& 0.530& 0.542&$1.17\!\times\!10^{-21}$&$2.95\!\times\!10^{  4}$\\
0.2300&&4.729&5.115 & 5.052& 0.459& 0.545&$1.42\!\times\!10^{-21}$&$4.12\!\times\!10^{  4}$\\
0.2200&&5.209&5.584 & 5.478& 0.428& 0.551&$1.54\!\times\!10^{-21}$&$4.73\!\times\!10^{  4}$\\
0.2175&&5.264& 5.626& 5.515& 0.424& 0.554&$1.56\!\times\!10^{-21}$&$4.79\!\times\!10^{  4}$\\
0.2150&&5.277&5.624 & 5.507& 0.423& 0.557&$1.55\!\times\!10^{-21}$&$4.80\!\times\!10^{  4}$\\
0.2100&&5.173& 5.503& 5.381& 0.428& 0.563&$1.52\!\times\!10^{-21}$&$4.64\!\times\!10^{  4}$\\
0.2000&&4.698& 5.052& 4.926& 0.452& 0.576&$1.39\!\times\!10^{-21}$&$4.07\!\times\!10^{  4}$\\
0.1900&&4.322&4.705 & 4.542& 0.466& 0.594&$1.28\!\times\!10^{-21}$&$3.66\!\times\!10^{  4}$\\
0.1800&&4.164&4.546 & 4.340& 0.457& 0.614&$1.22\!\times\!10^{-21}$&$3.56\!\times\!10^{  4}$\\
0.1600&&4.232& 4.585& 4.346& 0.403& 0.653&$1.23\!\times\!10^{-21}$&$3.92\!\times\!10^{  4}$\\
0.1400&&4.692&5.057 & 4.838& 0.360& 0.678&$1.36\!\times\!10^{-21}$&$4.67\!\times\!10^{  4}$\\
0.1200&&5.928&6.234 & 5.787& 0.321& 0.675&$1.63\!\times\!10^{-21}$&$5.93\!\times\!10^{  4}$\\
0.1000&&& & 7.492& 0.273& 0.649&$2.11\!\times\!10^{-21}$&$8.23\!\times\!10^{  4}$\\
\botrule
\end{tabular}
\end{center}
$^\www$ Tabulated data is available at
http://www.astro.princeton.edu/$\sim$draine/dust/dust.html\\
$^\ccm$ \citet{CCM89} extinction fit for $R_V=3.1$ \\
$^\fmod$ \citet{Fi99} extinction fit for $R_V=3.1$\\
$^\mmm$ $R_V=3.1$ Milky Way dust model of \citet{WD01a} 
but with abundances reduced by
0.93, and using optical constants from \citet{Dr03b}\\
$^\aaa$ $\lambda_{\rm eff}$ for $JHKLL^\prime M$ from 
	\citet{BB88};
	$R_JI_J$ from
	\citet{Jo65};
	Cousins $UBVR_CI_C$ from 
	\citet{FIG96};
	SDSS $ugriz$ from Gunn (2002, private communication).\\
$^\bbb$ $A_\lambda=(2.5/\ln10) C_{\rm ext}(\lambda) N_{\rm H} =
	1.086 C_{\rm ext}(\lambda) N_{\rm H}$\\
$^\ccc$ $\kappa_{\rm abs}(\lambda)=$ absorption cross section per unit
	dust mass.  For this model, $M_{\rm H}/M_{\rm dust} =90.$

}
\end{table}
\renewcommand \aaa {a}
\begin{table}
\caption{Absorption and Scattering$^\aaa$ for $\lambda > 5\micron$}
\label{tab:model_fir}
{\footnotesize
\begin{center}
\begin{tabular}{@{}ccccccc@{}}
\toprule
$\lambda$&$A_\lambda/A_{I_C}$&albedo&$\langle\cos(\theta)\rangle$&$C_{\rm ext}^{\ \aaa}$&
$\kappa_{\rm abs}^{\ \aaa}$\\
$\mu$m&model&&&cm$^2$/H&cm$^2$/g\\
\toprule
 1000.& .00002&.00000&-.004&$5.48\!\times\!10^{-27}$&$2.94\!\times\!10^{ -1}$\\
 850.0& .00003&.00000&-.005&$7.15\!\times\!10^{-27}$&$3.83\!\times\!10^{ -1}$\\
 350.0& .00013&.00000&-.005&$3.58\!\times\!10^{-26}$&$1.92\!\times\!10^{  0}$\\
 200.0& .00042&.00001&-.004&$1.19\!\times\!10^{-25}$&$6.37\!\times\!10^{  0}$\\
 140.0& .00092&.00001& .000&$2.59\!\times\!10^{-25}$&$1.39\!\times\!10^{  1}$\\
 100.0& .00180&.00003& .007&$5.07\!\times\!10^{-25}$&$2.71\!\times\!10^{  1}$\\
 60.00& .0058&.00007& .000&$1.62\!\times\!10^{-24}$&$8.70\!\times\!10^{  1}$\\
 40.00& .0156&.00014&-.016&$4.40\!\times\!10^{-24}$&$2.36\!\times\!10^{  2}$\\
 25.00& .0317&.00031&-.025&$8.95\!\times\!10^{-24}$&$4.79\!\times\!10^{  2}$\\
 18.00& .0514&.00066&-.026&$1.45\!\times\!10^{-23}$&$7.76\!\times\!10^{  2}$\\
 14.00& .0327&.00283&-.029&$9.22\!\times\!10^{-24}$&$4.92\!\times\!10^{  2}$\\
 12.00& .0622&.00301&-.029&$1.75\!\times\!10^{-23}$&$9.37\!\times\!10^{  2}$\\
 9.700& 0.140& .003&-.036&$3.94\!\times\!10^{-23}$&$2.10\!\times\!10^{  3}$\\
 8.000& .0491& .017&-.050&$1.38\!\times\!10^{-23}$&$7.29\!\times\!10^{  2}$\\
 7.000& .0270& .055&-.053&$7.60\!\times\!10^{-24}$&$3.85\!\times\!10^{  2}$\\
 6.000& .0329& .088&-.052&$9.27\!\times\!10^{-24}$&$4.53\!\times\!10^{  2}$\\
\hline
\botrule
\end{tabular}
\end{center}
}
$^\aaa$ See Table \ref{tab:model_IRoptuv}
\end{table}
\renewcommand \aaa {a}
\renewcommand \bbb {b}
\renewcommand \ccc {c}
\begin{table}
\caption{Absorption and Scattering$^\aaa$ for $\lambda < 0.1\micron$}
\label{tab:model_euvxray}
{\footnotesize
\begin{center}
\begin{tabular}{@{}cccccccc@{}}
\toprule
$\lambda$&$h\nu$&$A_\lambda/A_{I_C}$&albedo&$\langle\cos(\theta)\rangle$&$C_{\rm ext}^{\ \aaa}$&
$\kappa_{\rm abs}^{\ \aaa}$\\
$\mu$m&eV&model&&&cm$^2$/H&cm$^2$/g\\
\toprule
.09120&13.6&8.256& 0.248& 0.658&$2.33\!\times\!10^{-21}$&$9.37\!\times\!10^{  4}$\\
.04881&25.4 & 5.574& 0.291& 0.813&$1.57\!\times\!10^{-21}$&$5.97\!\times\!10^{  4}$\\
.02339&53.0 & 3.355& 0.366& 0.953&$9.46\!\times\!10^{-22}$&$3.21\!\times\!10^{  4}$\\
.02279&54.4& 3.577& 0.369& 0.949&$1.01\!\times\!10^{-21}$&$3.41\!\times\!10^{  4}$\\
.02214&56.0& 3.469& 0.393& 0.950&$9.78\!\times\!10^{-22}$&$3.18\!\times\!10^{  4}$\\
.01240&100 & 2.564& 0.452& 0.988&$7.23\!\times\!10^{-22}$&$2.12\!\times\!10^{  4}$\\
.01170&106 &  2.908& 0.432& 0.985&$8.20\!\times\!10^{-22}$&$2.49\!\times\!10^{  4}$\\
.006199&200 & 2.000& 0.578& 0.997&$5.64\!\times\!10^{-22}$&$1.27\!\times\!10^{  4}$\\
.004397&282 & 1.426& 0.619& 0.999&$4.02\!\times\!10^{-22}$&$8.20\!\times\!10^{  3}$\\
.004350&285 & 2.355& 0.452& 0.998&$6.64\!\times\!10^{-22}$&$1.95\!\times\!10^{  4}$\\
.004133&300 & 1.812& 0.517& 0.999&$5.11\!\times\!10^{-22}$&$1.32\!\times\!10^{  4}$\\
.002480&500 & 0.971& 0.632& 1.000&$2.74\!\times\!10^{-22}$&$5.40\!\times\!10^{  3}$\\
.002317&535 & 0.691& 0.386& 1.000&$1.95\!\times\!10^{-22}$&$6.40\!\times\!10^{  3}$\\
.002305&538 & 1.325& 0.355& 1.000&$3.74\!\times\!10^{-22}$&$1.29\!\times\!10^{  4}$\\
.002296&540 & 1.138& 0.453& 1.000&$3.21\!\times\!10^{-22}$&$9.40\!\times\!10^{  3}$\\
.002254&550 & 1.056& 0.478& 1.000&$2.98\!\times\!10^{-22}$&$8.32\!\times\!10^{  3}$\\
.001759&705 & 0.575& 0.433& 1.000&$1.62\!\times\!10^{-22}$&$4.92\!\times\!10^{  3}$\\
.001751&708 & 1.219& 0.353& 1.000&$3.44\!\times\!10^{-22}$&$1.19\!\times\!10^{  4}$\\
.001734&715 & 0.799& 0.515& 1.000&$2.25\!\times\!10^{-22}$&$5.84\!\times\!10^{  3}$\\
.001720&721 & 0.941& 0.432& 1.000&$2.65\!\times\!10^{-22}$&$8.07\!\times\!10^{  3}$\\
.001710&725 &  0.845& 0.503& 1.000&$2.38\!\times\!10^{-22}$&$6.33\!\times\!10^{  3}$\\
.001550&800 & 0.751& 0.514& 1.000&$2.12\!\times\!10^{-22}$&$5.51\!\times\!10^{  3}$\\
.001240&1000 & 0.563& 0.564& 1.000&$1.59\!\times\!10^{-22}$&$3.71\!\times\!10^{  3}$\\
.000620&2000 & 0.181& 0.584& 1.000&$5.10\!\times\!10^{-23}$&$1.14\!\times\!10^{  3}$\\
\botrule
\end{tabular}
\end{center}
}
$^\aaa$ see Table \ref{tab:model_IRoptuv}
\end{table}


\section{SCATTERING
	\label{sec:scattering}}
\subsection{Optical and UV}

Observations of light reflected from dust grains provides another test
of dust grain models. 
Individual reflection nebulae are bright, but comparisons with models
are uncertain because the scattering geometry is in general poorly-determined:
the illuminating star could be in front of the dust cloud (in which case
the typical scattering angle would be large) or it could be embedded within
it (in which case forward scattering by the dust grains between us and
the star would dominate).

The ``diffuse galactic light'' (DGL) consists of starlight scattered off the
dust in the diffuse ISM, illuminated by the general
interstellar starlight radiation field.  Since we think we have independent
knowledge of the spatial distribution of both stars and dust, studies
of dust properties using the DGL do not suffer from
geometric ambiguity.
However, the DGL is faint (and must be distinguished
from the direct light from faint stars), and attempts to infer the
grain albedo and scattering asymmetry factor $g\equiv\langle\cos\theta\rangle$
often find that there can be a one-parameter family of (albedo,$g$) values
which are consistent with the observations.

Figs.\ \ref{fig:albedo} and \ref{fig:g} show various observational estimates
of albedo and $g$ as a function of wavelength for interstellar dust,
based on observations of the DGL.
Determinations of grain scattering properties
in the reflection nebulae 
NGC~7023
\citep{WBP92},
IC435
\citep{CBG95},
NGC~2023 
\citep{BMF02},
and the Pleiades
\citep{GN03}
are also shown.
\begin{enumerate}
\item There does appear to be evidence for a decrease in albedo from the
	optical to the vacuum UV, although
	the results for NGC~7023 and IC435 at $\lambda^{-1}>5\micron^{-1}$
	do not conform.
\item The data are consistent with (but do not require)
	a rise in the scattering asymmetry
	factor $g$ from optical to UV.
\item The data in Figs.\ \ref{fig:albedo} and \ref{fig:g} appear to be
	consistent with the albedo and $g$ calculated for the
	carbonaceous-silicate grain model.
\end{enumerate}
The scatter in the results of different studies 
[e.g., the albedo and $g$
near $\lambda\approx1600\Angstrom$ 
($\lambda^{-1}\approx6.25\micron^{-1}$)
in Figs.\ \ref{fig:albedo} and \ref{fig:g}]
serves as a warning that determining the albedo is model-dependent,
requires accurate photometry, and, in the case of photodissociation
regions like NGC~7023, NGC~2023, or IC~435, may require subtraction
of fluorescent emission from H$_2$.

\subsection{X-Ray Scattering}

Interstellar grains scatter X-rays through small scattering angles, and
as a result X-ray point sources can appear to be surrounded by a diffuse
``halo'' of X-rays scattered by dust grains between us and the source
\citep{Ov65,Ha70,Mar70}.
Measurements of such halos provide a quantitative test of
interstellar grain models
\citep{Ca83,MG86,MTK90,ML91,CWN94,WCD94,MCF95,PK96,SD98,WSD01,SES02}.

ROSAT observations of Nova Cygni 1992 provide the best observational
data at this time.  
\citet{MCF95}
concluded that the observed
halo on day 291 (after optical maximum) 
was best explained if the larger ($a\geq0.1\micron$)
interstellar grains had a ``fluffy''
morphology, with a void fraction $\geq25\%$ -- if the grains were
compact the scattering halo would be stronger than observed.
\citet{SD98}
argued that \citet{MCF95} had overestimated
the scattering due to their use of the Rayleigh-Gans approximation,
and that in fact the observed X-ray halo was consistent with
compact grains.
The same data was reconsidered by 
\citet{WSD01},
who concluded that
the observed halo intensity at small scattering angles required a
substantial population of large ($a\geq1\micron$) grains.

\citet{DT03}
have revisited this question.
Using data for 9 epochs,
they find that the \citet{WD01a} PAH/graphite/silicate grain model -- with
no adjustment to the size distribution -- 
is consistent with the observed X-ray halo, for 
50$^{\prime\prime}$--3000$^{\prime\prime}$ halo angles.
Observations with ROSAT and Chandra of other sources at energies between
0.7 and 4 keV also appear to be in general agreement with the scattering
expected for this grain model \citep{Dr03b}.

\section{INFRARED EMISSION
	\label{sec:IRemission}}

\begin{figure}[htb]
\begin{center}
\epsfig{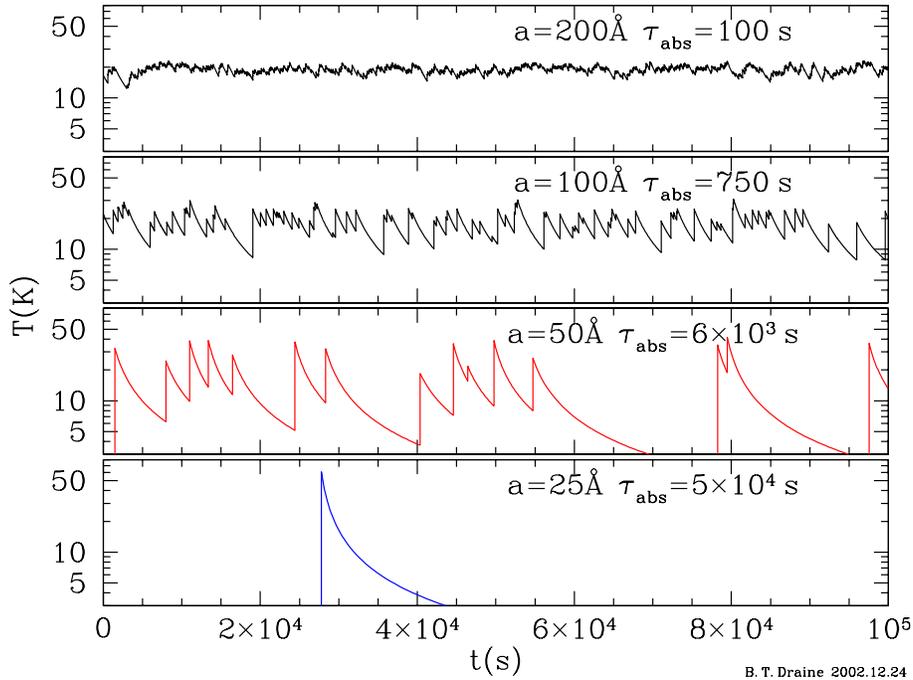}
\vspace*{-0.3cm}
\end{center}
%
%
%
\caption{\footnotesize
	\label{fig:Tfluc}
	A day in the life of 4 carbonaceous grains, heated by the
	local interstellar radiation field.
	$\tau_{\rm abs}$ is the mean time between photon absorptions
	\citep{DL01}.
	}
\end{figure}

Heating of interstellar dust grains is primarily by absorption of starlight
(collisional heating dominates only in dense regions in dark clouds -- where
the starlight intensity has been severely attenuated -- or in dense, hot,
shocked gas).
Since the heating by starlight photons is quantized and stochastic, the
temperature of a dust grain is time-dependent.
Figure \ref{fig:Tfluc} shows the temperature histories of 4 dust grains,
heated by the average starlight background, and cooled by emission of
infrared photons,
over a time span of about 1 day.
It is apparent that for grains heated by the average starlight background,
grains with radii $a\gtsim200\Angstrom$ can be approximated as having
a steady temperature.  Grains with radii $a\ltsim50\Angstrom$, however,
undergo very large temperature excursions, and the notion of ``average
temperature'' is not applicable.
Most of the infrared power radiated by such small grains occurs during brief
intervals following photon arrivals, when the grain temperature is close to
the peak.

In order to calculate the time-averaged emission spectrum for such small
grains, one must calculate the energy distribution function $dP/dE$,
for grains of a particular size, where $dP$ is the probability that a grain
will have vibrational energy in interval $(E,E+dE)$.
To solve for $dP/dE$, one must first calculate the specific heat for the
grain, and the cross sections for absorption and emission of photons.
Monte-Carlo simulations can be used to obtain $dP/dE$, but it is more
efficient to solve for the steady-state $dP/dE$ 
\citep{GD89,DL01}.

The notion of ``temperature'' is somewhat problematic for very small grains, 
but one can define the instantaneous vibrational temperature as being
equal to the temperature $T(E)$ at which the expectation value of the
energy would be equal to the actual grain energy; this is the temperature
plotted in Figure \ref{fig:Tfluc}.
\citet{DL01} have shown that a ``thermal'' estimate for
the grain emissivity is a good approximation, even for
grains containing as few as $\sim30$ atoms.

\begin{figure}[htb]
\begin{center}
\epsfig{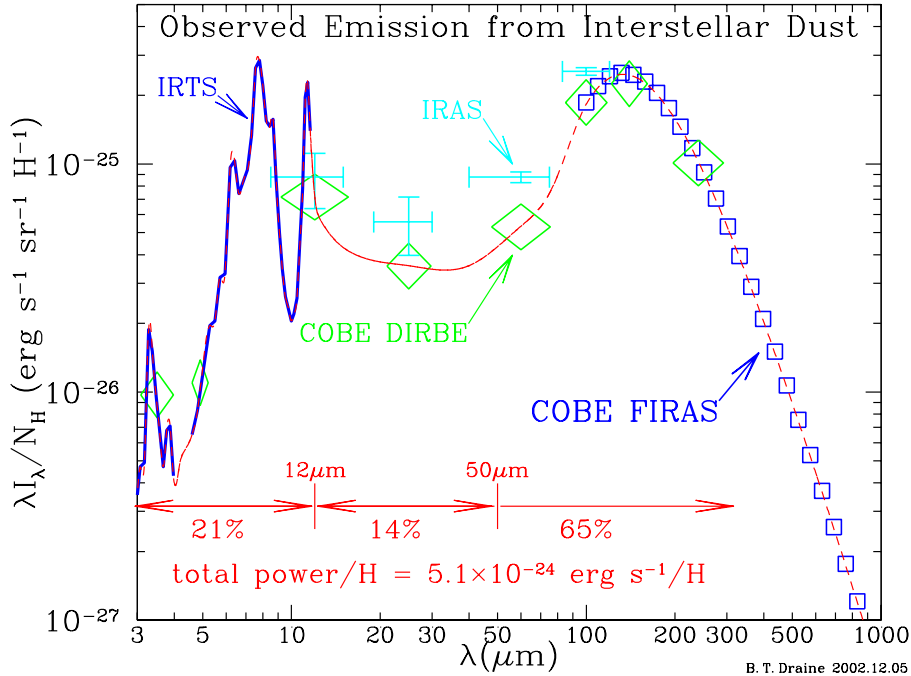}
\vspace*{-0.3cm}
\end{center}
%
%
%
\caption{\footnotesize
	\label{fig:fir}
	Observed emission from diffuse interstellar dust.
	Crosses: IRAS 
	\citep{BP88};
	Squares: COBE-FIRAS 
	\citep{FDS99};
	Diamonds: COBE-DIRBE 
	\citep{AOW98};
	Heavy Curve: IRTS 
	\citep{OYT96, TMM96}.
	The total power $\sim5.1\times10^{-24}\erg\s^{-1}/{\rm H}$
	is estimated from the interpolated broken line.
	}
\end{figure}

\begin{figure}[htb]
\begin{center}
\epsfig{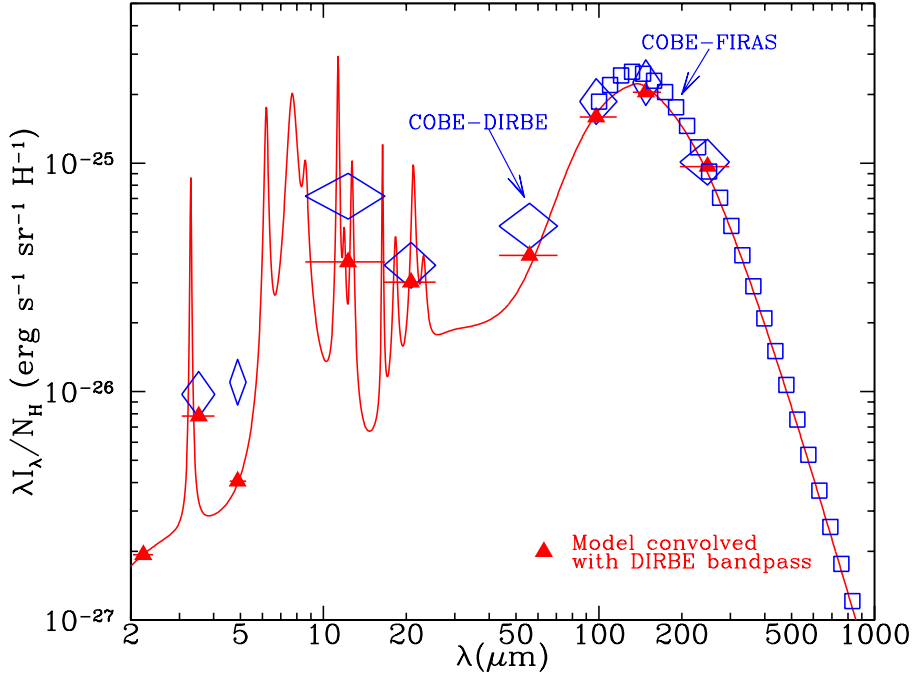}
\vspace*{-0.3cm}
\end{center}
%
%
%
\caption{\footnotesize
	\label{fig:hgl}
	Solid line: infrared emission calculated for 
	WD01 $R_V=3.1$ Milky Way dust model (but with abundances
	reduced by factor 0.93)
	for grains heated by diffuse starlight (Li \& Draine 2001b).
	Squares: COBE-FIRAS 
	\citep{FDS99}
	Diamonds: COBE-DIRBE 
	\citep{AOW98}.
	}
\end{figure}

The infrared and far-infrared glow from interstellar dust can be measured,
and a dust model
must be able to reproduce the observed intensity and spectrum.
The diffuse ISM at galactic latitudes $|b|\gtsim20^\circ$
provides a particularly clean test,
because we have good estimates for the intensity of the starlight
heating the dust, and 
the observed 21~cm emission provides accurate
gas column densities which, for an assumed dust-to-gas ratio,
determine the column density of dust.
Although the surface brightness of the infrared emission is low, it has
been detected by IRAS and COBE after averaging over large areas of sky.
Figure \ref{fig:fir} shows the observed emissivity per H nucleon, based
on COBE-FIRAS and COBE-DIRBE photometry.
3.0-4.5$\micron$ and 4.5-11.3$\micron$
spectra, measured at $b\approx0$ by
the IRTS satellite
have been added to the plot, normalized to agree with the DIRBE
photometry in the 3.5, 5.0, and 12$\micron$ bands.

In this dust model, 
approximately 20\% of the absorbed starlight energy is reradiated at
$\lambda<12\micron$ by PAHs.
The same grain model, applied to reflection nebulae,
can acount for the observed dependence of dust emission
spectrum on temperature of illuminating star 
\citep{LD02b}.
With appropriate changes to the grain mixture, the model can also
account for the observed IR emission from the SMC 
\citep{LD02c}.

Approximately two thirds 
of the radiated power is at $\lambda \gtsim 50\micron$ -- this
is emission from $a\gtsim .01\micron$ 
grains that are maintained at a nearly steady temperature
$\sim15-20\K$ by starlight.
The remaining $\sim$ one third of the radiated power is from grains with radii 
$a\ltsim50\Angstrom$ that are cooling
following ``temperature spikes'' resulting from absorption of individual
starlight photons.
We can immediately see the following:
({\it A}) The grain population must contain enough
$a\ltsim50\Angstrom$ grains to account for $\sim$ one third 
of the absorption of
starlight.
({\it B}) The resemblance of the 3--12$\micron$ emission spectrum from
the diffuse ISM to the emission spectrum expected from polycyclic
aromatic hydrocarbon (PAH) molecules, radicals, and ions implies that
a large fraction of the smallest grains must be PAHs.
Figure \ref{fig:hgl} shows the emission calculated for the
PAH/carbon/silicate dust model illuminated by the average interstellar
radiation field spectrum.

The far-infrared emission is a good tracer of interstellar dust.
The IRAS and DIRBE all-sky maps can be used to infer the dust column
density 
\citep{SFD98}.

\section{MICROWAVE EMISSION
	\label{sec:microwave}}

\subsection{Spinning Grains}

Theoretical models to account for the observed infrared emission require
large numbers of extremely small dust grains that, following absorption
of starlight photons, radiate in their optically-active vibrational modes.
In addition to vibrating, these grains will
rotate, with rotation rates reaching tens of GHz.
If they have electric dipole moments, there will be rotational emission from
these grains at microwave frequencies.
Various workers have considered microwave emission from spinning grains
\citep{Er57,HW70,FD94}.
The rotational emission from this grain population has been estimated
by \citet{DL98a,DL98b},
who showed that it could account for
dust-correlated microwave emission that was discovered by 
\citet{KBB96a,KBB96b}
in studies of the angular structure of the cosmic microwave background.

\begin{figure}[htb]
\begin{center}
\epsfig{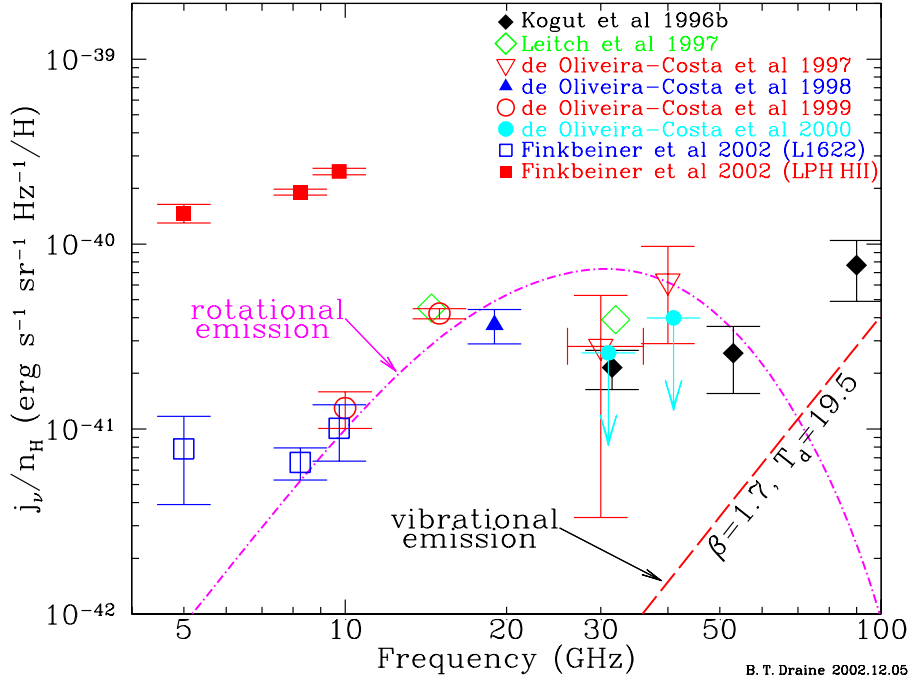}
\vspace*{-0.3cm}
\end{center}
%
%
%
\caption{\footnotesize
	\label{fig:microwave}
	Dust-correlated microwave emission, expressed as microwave
	surface brightness per unit column density of hydrogen.
	Curves: theoretical estimate
	of \citet{DL98a,DL98b}.
	Symbols: observational determinations of ``anomalous microwave
	emission''.
	}
\end{figure}

The theoretical estimate for the rotational emission, plus the low-frequency
tail of the vibrational emission, is shown in Figure \ref{fig:microwave}.
Figure \ref{fig:microwave} also shows results from a number of 
observational studies --
some over large areas of the sky, others over
smaller regions.
\citet{FSF02}
detect dust-correlated
microwave emission in pointed observations of a dark cloud (L1622) and
toward LPH 201.663+1.643, a diffuse HII region behind heavy obscuration.
Most recently, 
\citet{FLM03}
use 8 and 14 GHz surveys to
map the anomalous microwave emission from the Galactic plane.

With the exception of the observations toward LPH 201.663+1.643, the
observed microwave emission appears to be generally consistent with
the theoretical prediction for emission from
spinning dust grains.

The observed microwave emission from LPH 201.663+1.643 is, however, an
order of magnitude stronger than would be expected from the observed
column of dust.  While the rotational emission from spinning dust grains
does depend on environment 
\citet{DL98b}, it does not
appear possible to explain the strong emission observed in this direction
using reasonable variations in environmental conditions.

\subsection{Magnetic Grains?}

Although rotational emission from spinning grains is a natural prediction
for a dust model with large numbers of small grains, other emission 
mechanisms are also possible, including thermal magnetic dipole
radiation from magnetic grains 
\citep{DL99}.
We know that Fe contributes a significant fraction of the interstellar
dust mass; if some of this Fe is in ferrimagnetic or 
ferromagnetic compounds, grains containing these compounds would emit
magnetic dipole radiation.  The idea is that the magnetization of the
magnetic domains in the grain would fluctuate around the minimum-energy
state; these fluctuations would be at GHz frequencies, and the fluctuating
magnetization would generate electromagnetic radiation.
If an appreciable fraction of the Fe in interstellar grains were in
magnetite Fe$_3$O$_4$, it is estimated that 
the resulting thermal magnetic dipole emission
would be comparable to the observed microwave emission.
The power radiated per grain is simply proportional to the mass
of magnetic material in that grain, and proportional to the grain
temperature.  
Since it is presumed that most of
the magnetic material would be in large grains, it would be expected that
the large grains would account for the bulk of the magnetic dipole emission.

It is therefore possible that the observed microwave emission may contain
appreciable contributions from both (ultrasmall) spinning grains and 
(large) magnetic grains.

\section{ABUNDANCE ISSUES
	\label{sec:abundances}}

Dust population models that are constructed to reproduce the observed
wavelength-dependent extinction are found to imply substantial abundances
of elements in
grain material -- approaching or exceeding the abundances believed
to be appropriate to interstellar matter.
\citet{KM96}
referred to the ``C/H crisis'', and 
\citet{Ma96}
has argued that the problem may be even more severe for
Mg, Si, and Fe required for silicate grains.

Whether there is a problem or not depends on the adopted values for
total interstellar abundances.  
\citet{Ma96}
took the position that
interstellar abundances should be taken to be $\sim$70\% of solar.
However, recent changes in inferred solar abundances for C and O, and
comparison of photospheric abundances in F and G stars with those in
B stars, led 
\citet{SM01}
to reject the hypothesis that interstellar abundances are appreciably subsolar.
Sofia \& Meyer estimate the dust-phase abundances on the sightline 
to $\zeta$Oph if interstellar abundances are taken to be the same
as photospheric abundances in F and G stars.
From these abundances, we can construct carbon grains with
volume = $2.05\times10^{-27}\cm^3/{\rm H}$ (for an assumed 
$\rho=2.2\g\cm^{-3}$), and silicate/SiO$_2$ grains with
volume $2.90\times10^{-27}\cm^3/{\rm H}$ (for an assumed $\rho=3.5\g\cm^{-3}$).

The provisional grain model of \S\ref{sec:prov_model} -- the WD01
grain model with abundances reduced by a factor 0.93 -- requires
a carbonaceous grain volume $V_{\rm carb}=2.10\times10^{-27}\cm^3/{\rm H}$,
and a silicate grain volume $V_{\rm sil}=3.68\times10^{-27}\cm^3/{\rm H}$:
thus we see that the carbon abundance is in line with observations, but
the provisional grain model requires $\sim$26\% more Mg, Si, and Fe than
appears to be available.

Is this 26\% discrepancy a serious problem?
Consider the following:
\begin{enumerate}
\item 6 of the 14 sightlines in Figure \ref{fig:extperH}
have $A_I/N_{\rm H}$ differing by more than 30\% from the
nominal value.  This may represent measurement errors, but suggests
real variations in $A_I/N_{\rm H}$.
Since it is not clear how the sample of sightlines is constructed, it
is not apparent that the average value of $N_{\rm H}/A_I$ for the sample
is the same as the overall dust-to-gas ratio for the ISM.
\item The dust model assumes the grains to be spheres.
Real grains are nonspherical.
This will tend to provide somewhat more extinction per unit mass of
grain material; this could increase the visual extinction per unit
grain mass by $\sim10-20$\%.
\item The dust model assumes the grains to be compact, but
real grains may contain voids.
This could conceivably increase the visual extinction per unit grain
mass by $\sim10\%$.
\item The dust model uses ``bulk'' optical constants and Mie theory.
The outer few monolayers of grain material, however, will have electronic
structure (and therefore optical properties) differing from ``bulk'' material,
so that the model calculations should not be considered to
be precise even if the grains were actually spheres and we 
had the correct bulk optical constants.
\end{enumerate}
At the present stage of observational and theoretical knowledge, we do not
consider the apparent ``abundance crisis'' to be sufficient grounds to
reject our provisional grain model, but new data could change 
this conclusion.

\section{GRAIN ALIGNMENT
	\label{sec:alignment}}

The polarization of starlight (\S\ref{sec:pol}) was discovered over 
50 years ago 
\citep{Ha49,Hi49},
and was immediately recognized as being due to aligned dust grains.
The mechanism responsible for the alignment has
been sought by theorists for the past half century.
Two separate alignments are involved: (1) alignment of the
grain's principal axis of largest moment of inertia with the grain's
angular momentum ${\bf J}$, and 
(2) alignment of ${\bf J}$ with the galactic magnetic field.

The alignment processes turn out to involve quite subtle
physics [for an introduction, see 
\citet{Dr03a}].
The paramagnetic dissipation process proposed by 
\citet{DG51}
would be able to gradually align grains,
but recent work shows that radiative torques
due to starlight
can be much stronger than the Davis-Greenstein torque;
the starlight torques can affect both the rate of grain rotation -- driving
grains to suprathermal rotation rates -- as well as the direction of
the grain angular momentum
\citep{DW96,DW97}.
The electronic and nuclear paramagnetism of the grain couple the rotational
degrees of freedom to the vibrational degrees of freedom, with important
consequences for
the rotational dynamics of a tumbling grain (Lazarian \& Draine 1999a,b).

H$_2$ formation on grain surfaces is thought to take place at a limited number
of reaction sites, and the recoil from nascent H$_2$ provides a
systematic torque on the grain that is fixed in body coordinates
\citep{Pu79},
and that can drive a grain to suprathermal rotation rates.
When the grain rotational kinetic energy is not too large compared to
$kT_d$, where $T_d$ is the grain temperature, the process of
``thermal flipping'' 
\citep{LD99a}
can cause the H$_2$ formation to time-average to zero, so that suprathermal
rotation does not take place -- a process termed ``thermal trapping''.
Radiative torques due to starlight, however, are not fixed in grain body
coordinates and therefore do not time-average to zero even when thermal 
flipping is rapid, so that grains subject to starlight torques are
immune to ``thermal trapping'', and are expected to rotate suprathermally
most of the time.
Starlight torques are important only for grains with radii $a\gtsim0.1\micron$.
This may explain the
fact
that grains with radii $a\gtsim0.2\micron$ appear to be much
more strongly aligned than grains with radii $a\ltsim0.05\micron$
\citep{KM95}.

The theory of grain alignment is complicated by the need to describe
the dynamics of tumbling grains 
during intervals when the grain rotational kinetic
energy is of order $kT_{\rm gas}$, including the effects of
thermal flipping 
\citep{WD03}.
Progress is being made,
and quantitative calculations of grain alignment for model grains should
be possible within the next year or two -- this should allow us to
``predict'' the wavelength-dependent polarization for a grain model once
a geometric asymmetry is assumed for the grains.
At that time we will learn whether our current picture of grain dynamics
is consistent with observations.

The very smallest grains spin with GHz rotation rates and emit electric
dipole radiation in the microwave (\S\ref{sec:microwave}).
If these grains are aligned with the magnetic field, the rotational
emission will be polarized.
It appears, however, that these grains will be only minimally aligned,
with degrees of alignment of only a few percent 
\citep{LD00}.
The Davis-Greenstein paramagnetic dissipation mechanism is suppressed
in these grains by the unavailability of vibrational modes with energies
$\hbar\omega_{\rm rot}$.

\section{EVOLUTION OF THE GRAIN POPULATION}

The interstellar grain population in a galaxy reflects the interplay
of many processes.

Are interstellar grains ``stardust''?
There is ample evidence that grains form in some stellar outflows --
we see infrared emission from dust
in outflows from red giants and carbon stars,
and in planetary nebulae.
In all of these cases it is believed that the bulk of the dust 
condensed out of initially dust-free gas from the stellar envelope.
The 10$\micron$ silicate feature is seen in outflows from oxygen-rich
stars, consistent with the expectation that silicates should be a major
condensate in cooling gas with O/C $>$ 1.
The 10$\micron$ feature is {\it not} seen in outflows from 
stars with C/O $>$ 1, 
where instead one sometimes sees a feature at 11.3$\micron$ that is
attributed to SiC 
\citep{TC74,BBF98},
consistent with the expected condensates from cooling gas with O/C $<$ 1.
Because isotopic anomalies are used to find them, presolar grains in
meteorites generally appear to be ``stardust'' -- a given grain was
formed in the outflow/ejecta from a single star with a particular isotopic
composition.

There is no doubt that dust is formed in many stellar outflows.
However, this does {\it not} imply that the bulk of the dust in the ISM
was produced in this way.

The problem is that interstellar grains are subject to sputtering 
\citep{DS79a}
and grain-grain collisions in interstellar shock waves, and
theoretical studies of grain destruction in the Milky Way find 
a mean ``residence time'' of an atom in a grain of order
$\tau_{\rm dest}\approx 3\times10^8\yr$
\citep{DS79b,JTH94}.

Now suppose that all Si atoms enter the ISM as constituents of
dust grains.  The total mass of the ISM is 
$M_{\rm ISM}\approx5\times10^{9}M_\sol$, and star formation takes place
at a rate $\dot{M}_\star\ltsim 5 M_\sol\yr^{-1}$.
Thus the mean residence time of a metal atom in the ISM is 
$\tau_{\rm ISM}=M_{\rm ISM}/\dot{M}_\star\approx10^9\yr$.
Only
a fraction $\tau_{\rm dest}/(\tau_{\rm dest}+\tau_{\rm ISM}) \approx 0.2$
of the Si atoms in the ISM would be found in the original stardust particle
in which they left a star.

However, when we observe gas-phase abundances in the ISM, we typically
find $\gtsim90\%$ of the Si atoms missing from the gas phase.
Unless we have grossly underestimated the grain destruction time
$\tau_{\rm dest}$, it must follow that {\it most of the Si atoms in
interstellar grains joined the grain in the ISM!}

Therefore {\it most interstellar dust is not stardust} -- the composition
of interstellar grains is determined by physical processes in the ISM.
The importance of interstellar processing is also evident from the large
variations seen in the extinction curve from one sightline to another --
these variations must be the result of interstellar effects, since
any given sightline would be expected to average over the stardust
produced by many individual stars.

The interstellar processes that alter grains include
\begin{itemize}
\item grain erosion by sputtering;
\item grain shattering in low-velocity grain-grain collisions;
\item grain vaporization in high-velocity grain-grain collisions;
\item grain coagulation;
\item accretion of impinging atoms and molecules on grain surfaces;
\item chemical reaction of impinging atoms and radicals with grain surfaces;
\item photolysis of grain material by UV starlight;
\item alteration of grain material by cosmic ray/X-ray irradiation;
\item photodesorption of atoms and molecules from grain surfaces.
\end{itemize}
In addition, the local grain mixture can be modified by
size- and composition-dependent drift of dust through the gas, driven
by anisotropic starlight: relative drift speeds $\gtsim0.1\kms$ can
displace large grains from small grains by $\sim$few pc in $10^7\yr$
\citep{WD01b}.

Understanding the balance among these different processes requires not only
an understanding of the microscopic physics but also a quantitative model
for the transport of grains from one ``phase'' of the ISM to another.
Because grain destruction appears to be dominated by processes associated
with interstellar shock waves, the propagation of these shock waves through
the inhomogeneous ISM must be modeled.

The required modeling is complex, and we are not yet at a stage where
``a priori'' modeling of the evolution of the interstellar grain
population can be considered reliable.
However, the observed depletions can be used to constrain models of
the dynamic ISM:
in order for grains to be able to ``scavenge'' elements like
Si from the gas, 
there must be rapid ``cycling'' of matter between the diffuse
low-density phases (where most grain destruction takes place)
and denser regions (where densities are high enough for grain growth
to be effective), with cycling times as short as $\sim10^7\yr$
\citep{Dr90}.

\section{FUTURE DIRECTIONS}

Our present understanding of interstellar grains is built upon an
observational base.
New observational facilities can be expected to produce
important extensions to this observational base, possibly requiring
revisions to our models.
\begin{itemize}
\item SIRTF (Space InfraRed Telescope Facility) 
	will provide low and medium resolution 5--40$\micron$ spectra of
	reflection nebulae, 55--96$\micron$ spectral energy distributions,
	and imaging in 7 photometric bands from 3.5 to 160$\micron$.
	We can anticipate detection of new emission
	features in the 15--40$\micron$ region (where ISO had limited
	sensitivity), and studies of variations of dust emission features and
	spectra with environment.
\item Astro-F 
	\citep[][and references therein]{PSM02},
	scheduled for launch in 2004, will obtain
	all-sky maps in 4 bands between 
	50 and 200$\micron$, with improved spatial resolution and much
	better sensitivity than IRAS. 
	This will greatly improve our knowledge of
	the distribution of diffuse dust, the degree to which dust is
	correlated with H~I~21cm emission, and spatial variations in the 
	far-infrared band ratios.
	Astro-F may be able to provide all-sky imaging in 6--11$\micron$ 
	that would provide a map of PAH emission.
	Correlation of the PAH emission with 
	microwave emission would test the hypothesis that
	the microwave emission comes from spinning grains.
\item WMAP (Wilkinson Microwave Anisotropy Probe) 
	is providing maps of the microwave sky in 5 bands between 22 and 90
	GHz, with FWHM of 0.53$^\circ$ at 40 GHz.
	These maps will help characterize the spectrum of the
	microwave emission from dust, as well as variations in dust
	emission (intensity and spectrum) from one region to another
	\citep{BHHN03}.
	It will be of great interest to cross-correlate the all-sky microwave
	maps with infrared and far-infrared maps from IRAS, COBE, and,
	when available, Astro-F.
\item Planck/Surveyor, scheduled for launch in 2007, 
	will provide all-sky submm maps	that will determine
	the diffuse dust emission spectrum from 30 - 857 GHz
	(1 cm -- 350$\micron$).
\item Herschel, scheduled for launch together with Planck, 
	will be capable of sensitive
	broadband diffraction-limited imaging at 60--700$\micron$, extending
	beyond the 160$\micron$ cutoff of SIRTF.
	The longer wavelength coverage will provide improved
	determinations of the emission from cool dust, and will allow a census
	of dust masses in nearby galaxies.
\item The Stardust spacecraft is expected to return to Earth in 2006.
	Its aerogel panels will have interstellar grain material at
	impact sites
	and (we hope) intact cometary dust particles.
\item SOFIA (Stratospheric Observatory for Infrared Astronomy)
	will provide higher angular resolution dust emission imaging
	between 5 and 240$\micron$
\item Ground-based millimeter and submillimeter telescopes and arrays 
	(e.g, BIMA, JCMT, CSO, ALMA) can map thermal dust emission from
	dense regions, including polarimetric mapping.
\end{itemize}

Many problems await solution, some of which are highlighted here:
\begin{itemize}
\item The carrier of the 2175$\Angstrom$ feature remains unidentified.
	The PAHs required by the observed IR emission features
	are expected to produce strong absorption in this
	region, but this remains conjecture.  We require
	lab measurements of UV absorption by 
	PAH molecules, radicals, and ions containing 20--$10^3$ C atoms
	for comparison with the observed 2175\AA\ profile.
	
\item The nature of the ERE carrier remains an open question.
	The high intensities reported for diffuse clouds imply that the
	ERE carrier must be a major grain constituent (\S\ref{sec:ERE}).
	Further observations of the ERE are needed, particularly to confirm
	the intensity of ERE from diffuse clouds.  
	Further spectroscopy of ERE emission
	from bright regions,
	and laboratory studies of candidate ERE carriers,
	may lead to identification of the ERE
	carrier.

\item We should be able to identify at least some of the DIBs!
	Gas-phase spectroscopy of candidate molecules, radicals, and ions
	is needed.  When the right candidates are studied,
	positive spectroscopic identification
	of DIBs will follow.

\item PAH molecules, radicals, and ions appear to be a major part of
	the grain population in the Milky Way, but the origin of these
	PAHs is unknown.  Further observational study of spatial variations in
	PAH abundance (and perhaps spectral properties) may provide
	valuable clues.

\item We would like to understand the balance of processes (growth,
	coagulation, erosion, shattering) accounting for the size distribution
	of interstellar dust.  Observational determinations of regional
	variations in the size distribution will frame the theoretical
	attack on this problem.

\item Are there really separate populations of carbonaceous grains and 
	silicate grains?  If so, how does grain growth in the ISM
	maintain these separate populations?

\item Is the anomalous microwave emission due primarily to spinning
	dust grains?
	Is there a significant contribution 
	from magnetic dust materials (\S\ref{sec:microwave})?

\item Can we diagnose the composition of interstellar dust using X-ray
	absorption measurements(\S\ref{sec:xrayedge})?

\item The problem of interstellar grain alignment (\S\ref{sec:alignment})
	seems ripe for solution.
	Radiative torques due to starlight appear to be an important
	part of the alignment process.
	A solution to this problem may be in sight.

\end{itemize}

I thank D Finkbeiner,
D Gutkowicz-Krusin, A Li, and J Weingartner for helpful comments, 
and RH Lupton for 
availability of the SM software package.
This research was supported in part by NSF grant AST-9988126.


\end{document}